\documentclass[10pt,aps,prd,reprint,superscriptaddress,longbibliography,nofootinbib,floatfix]{revtex4-2}
\usepackage{mathrsfs,amsmath,amsthm,latexsym,amssymb,amsfonts,graphicx}
\usepackage{lmodern}
\usepackage[utf8]{inputenc}
\usepackage[T1]{fontenc}
\usepackage[colorlinks=true,linkcolor=blue,citecolor=blue,urlcolor=blue]{hyperref}
\usepackage{orcidlink}
\usepackage{booktabs}
\usepackage{bm}
\usepackage{microtype}
\allowdisplaybreaks

\newcommand{\UFPB}{Departamento de Física, Universidade Federal da Paraíba, Centro de Ciências Exatas e da Natureza, 58051-970, João Pessoa, Paraíba, Brazil}
\newcommand{\UCCB}{Programa de P\'os-Gradua\c c\~ao em F\'{\i}sica \& Coordena\c c\~ao do Curso de F\'{\i}sica - Bacharelado, Universidade Federal do Maranh\~{a}o, 65085-580 S\~{a}o Lu\'{\i}s, Maranh\~{a}o, Brazil}

\begin{document}
\baselineskip=12pt

\title{Shadow, acoustic redshift, and transfer observables of Lorentz-violating rotating acoustic black holes}

\author{Fernando M. Belchior\orcidlink{0009-0006-8675-7849}}
\email{fernandobelcks7@gmail.com}
\affiliation{\UFPB}
\author{Edilberto O. Silva\orcidlink{0000-0002-0297-5747}}
\email{edilberto.silva@ufma.br}
\affiliation{\UCCB}

\begin{abstract}
We develop an impact-parameter-resolved transfer analysis for the rotating acoustic black hole with Lorentz symmetry violation. The background is the $(2+1)$-dimensional Lorentz-violating draining-bathtub geometry, where the draining parameter $A$ fixes the sonic horizon, the circulation parameter $B$ controls rotation, and the Lorentz-breaking parameter $\alpha$ deforms the effective acoustic metric. We derive the null-ray equations, the critical-impact-parameter conditions, the acoustic shadow interval, and the redshift transfer factor. We then formulate an intensity-transfer prescription for thin rings and extended disks that accounts for source emissivity, emitter motion, finite source width, and detector convolution. The resulting observables form a hierarchy: the shadow width probes Lorentz-violating broadening, the shadow centroid traces rotation, the left-right acoustic-redshift asymmetry tests branch-dependent Doppler and frame-dragging effects, and the integrated flux asymmetry measures their imprint on the observed intensity. We also construct synthetic two-dimensional acoustic screen maps, showing that the $(2+1)$-dimensional capture interval is naturally represented as a vertical strip whose displacement and brightness imbalance encode the combined effects of $B$ and $\alpha$. We focus on the exterior-regular regime $\alpha\geq0$, with $\alpha=0$ retained as the Lorentz-symmetric benchmark.\\
{\bf Keywords:} acoustic black hole; analogue gravity; draining bathtub; acoustic shadow; redshift distribution; observed flux; ray tracing
\end{abstract}

\maketitle

\section{Introduction}

In the last few years, analog gravity models have provided an interesting testing ground for studying horizon physics by propagating waves in media. The central observation is that linear sound perturbations in a moving fluid obey a wave equation on an effective curved spacetime under suitable hydrodynamic assumptions. This is the idea behind the so-called acoustic black hole introduced by Unruh \cite{Unruh1981} and later developed by Visser with a more general effective-metric framework \cite{Visser1998}. These seminal studies have motivated a broad literature on sonic horizons, ergoregions, Hawking-like emission, mode conversion, and laboratory analogs of black-hole phenomena~\cite{Barcelo2011,Novello2002,Garay2000}. In this direction, experiments and analogue platforms have explored several of these ideas, including stimulated Hawking emission, analogue superradiance and correlated phonon emission~\cite{Philbin2008,Lahav2010,Weinfurtner2011,Steinhauer2016,Torres2017,Patrick2018,Munoz2019,Kolobov2021}. Ultracold-gas and Bose-Einstein-condensate platforms have also extended the scope of analogue gravity toward quantum simulation of curved spacetimes, including analogue cosmology, cosmological quasiparticle production, sonic de Sitter/Gibbons-Hawking physics, observer-dependent particle content, trans-Planckian dispersion effects and quantum backreaction beyond the purely linearized acoustic metric~\cite{FedichevFischer2004PRA,FischerSchutzhold2004,FedichevFischer2003,FedichevFischer2004PRD,ChaFischer2017,ChandranFischer2025,SchuetzholdUhlmannXuFischer2005,BaakRibeiroFischer2022,PalFischer2024}.

In this context, the rotating draining bathtub flow is one of the simplest geometries in which acoustic horizons and rotational frame-dragging coexist. Its radial draining parameter determines the sonic horizon, while its circulation parameter generates an ergoregion. Early work on irrotational vortex acoustics already showed that the acoustic geometry of a hydrodynamical vortex can produce phonon deflection analogous to gravitational lensing~\cite{FischerVisser2002}. The draining-bathtub geometry has since been used to analyze scattering, absorption, superradiance, quasinormal modes, and orbiting phenomena in analog black holes~\cite{Basak2003,Berti2004,Cardoso2004,Dolan2009,Oliveira2010,Dolan2012,Dolan2013,Berti2009}. In parallel, Lorentz-violating (LV) extensions of field theory and effective geometry provide a phenomenological way to test how symmetry-breaking corrections modify wave propagation~\cite{Colladay1997,Colladay1998,Kostelecky2004,Mattingly2005,Liberati2013}. Acoustic black holes in LV Abelian-Higgs-type models have therefore become a useful laboratory for studying how such corrections change superresonance, analog Aharonov-Bohm scattering, absorption, and quasinormal modes~\cite{Anacleto2011,Anacleto2012,Anacleto2019,Campos2026}.

The purpose of the present work is complementary to absorption and quasinormal-mode calculations. We ask how the rotating LV acoustic black hole appears in an impact-parameter-resolved acoustic image. The question is inspired by the logic of black-hole imaging: a shadow boundary is a geometric capture feature, but the observed brightness is a transfer observable. The received intensity depends not only on the critical null rays, but also on the source distribution, emitter velocity, redshift, caustic magnification and detector response~\cite{Synge1966,Bardeen1973,Cunningham1973,Luminet1979,Falcke2000,Cunha2018,Gralla2019,Gralla2020,Johnson2020,Perlick2022}. In relativistic imaging this separation between geometry and transfer is essential for interpreting Event Horizon Telescope observations~\cite{EHT2019I,EHT2019VI,EHT2022I,EHT2022VI,Narayan2019,Dexter2016}. It is also central in modern ray-tracing and transfer calculations involving hot-spot images, centroid motion, light curves, polarized covariant radiative transfer, and medium-induced modifications of black-hole shadows~\cite{HuangZhangGuoChen2024,HuangZhengGuoChen2024,ZhongHuYanGuoChen2021}. The same separation is useful in acoustic analogues.

Following the transfer-observable framework developed for the ordinary rotating draining bathtub~\cite{Ahmed2026Transfer}, we separate the calculation into three layers. The first layer is purely geometric, where the null-ray structure and the critical impact parameters define the acoustic shadow interval. The second layer is kinematic, in which the acoustic redshift depends on the emitter's velocity, the ray branch, and the Lorentz-violating deformation of the metric. The third layer is observational, where the measured intensity depends on the transfer exponent, source emissivity, source width, and detector convolution. This hierarchy avoids confusing a source-dependent brightness contrast with a source-independent capture boundary.

This separation is especially important in the LV case. The circulation parameter $B$ breaks the left-right symmetry of the image and shifts the shadow centroid. The LV parameter $\alpha$ changes the effective acoustic scale, the circumference function, the redshift normalization, and the critical capture interval. Consequently, the width, centroid, redshift asymmetry, and flux asymmetry carry different physical information. A robust inference should therefore fit the geometry first, then check whether the transferred redshift and brightness profiles are consistent with the same values of $\alpha$ and $B$.

The paper is organized as follows. Section~\ref{sec:geometry} introduces the Lorentz-violating rotating acoustic geometry. Section~\ref{sec:nullrays} derives the null-ray equations and the shadow interval. Section~\ref{sec:redshift} develops the acoustic-redshift formula. Section~\ref{sec:transfer} formulates the intensity-transfer prescription for rings and disks. Section~\ref{sec:2dscreen} constructs the synthetic two-dimensional acoustic screen maps. Section~\ref{sec:observables} defines the differential observables. Section~\ref{sec:discussion} discusses the physical interpretation, parameter degeneracies, and consistency checks. Section~\ref{sec:conclusion} summarizes the results.

\section{Lorentz-violating rotating acoustic geometry}
\label{sec:geometry}

In this section, we provide a short review of
the rotating acoustic black hole with Lorentz symmetry violation introduced in Ref.~\cite{Campos2026}. In the original convention, the line element is written as
\begin{equation}
 ds^2=-(1+\alpha)F(r)d\tau^2-2B\,d\phi\,d\tau
 +G^{-1}(r)dr^2
 +\gamma^2(r)d\phi^2 ,
 \label{eq:original_metric}
\end{equation}
where
\begin{equation}
 F(r)
 =
 1-\frac{A^2+B^2}{(1+\alpha)r^2},
 \qquad
 G(r)
 =
 1-\frac{A^2}{(1+\alpha)r^2},
 \label{eq:FG_functions}
\end{equation}
and
\begin{equation}
 \gamma(r)
 =
 r\left[
 1+\frac{2\alpha(A+B)}{r}
 \right]^{1/2}.
 \label{eq:gamma_function}
\end{equation}
The parameters have the following meaning. The constant $A$ is associated with the radial draining flow, $B$ is associated with the vortex circulation, and $\alpha$ controls the Lorentz-symmetry-breaking correction. In the limit $\alpha\to0$, Eq.~\eqref{eq:original_metric} reduces to the usual rotating draining-bathtub acoustic geometry.

\subsection*{Domain of validity}

The angular circumference function $\gamma^2(r) = r^2 + 2\alpha(A+B)r$ must remain non-negative throughout the exterior region $r>r_h$ to preserve a well-defined effective acoustic geometry. For $\alpha>0$, this condition is satisfied for all $r>0$ provided $A+B>0$, which holds for the physical draining-vortex configuration. For $\alpha<0$, however, $\gamma^2(r)$ vanishes at $r_0 = 2|\alpha|(A+B)$ and becomes negative for $r<r_0$. If $r_0<r_h$, the singular shell lies inside the horizon and does not affect exterior ray propagation. If $r_0\geq r_h$, the effective geometry breaks down in a region accessible to external acoustic rays, and the present framework no longer applies without modification. Accordingly, throughout this work we restrict to the regime $\alpha\geq0$, which is also the physically motivated branch in the absorption analysis of Ref.~\cite{Campos2026}. The limit $\alpha=0$ is always included as a check.

For the ray-tracing analysis, it is convenient to multiply the metric by an overall minus sign and use the mostly-minus convention. Since null rays are unchanged by an overall conformal sign, we write
\begin{equation}
 ds^2
 =
 {\cal H}(r)dt^2
 -\frac{dr^2}{{\cal G}(r)}
 +2B\,dt\,d\phi
 -\gamma^2(r)d\phi^2 ,
 \label{eq:metric_mostly_minus}
\end{equation}
where
\begin{equation}
 {\cal H}(r)
 =
 (1+\alpha)-\frac{A^2+B^2}{r^2},
 \qquad
 {\cal G}(r)
 =
 1-\frac{A^2}{(1+\alpha)r^2}.
 \label{eq:H_and_G}
\end{equation}
The nonzero metric components are therefore
\begin{align}
& g_{tt}={\cal H}(r),
 \qquad
 g_{t\phi}=B,\\
 & g_{rr}=-{\cal G}^{-1}(r),
 \qquad
 g_{\phi\phi}=-\gamma^2(r).
 \label{eq:metric_components}
\end{align}

With that, the acoustic horizon is determined by ${\cal G}(r_h)=0$, giving
\begin{equation}
 r_h=\frac{A}{\sqrt{1+\alpha}}.
 \label{eq:horizon}
\end{equation}
Meanwhile, the ergosurface is determined by ${\cal H}(r_e)=0$, giving
\begin{equation}
 r_e=\sqrt{\frac{A^2+B^2}{1+\alpha}} .
 \label{eq:ergosurface}
\end{equation}
At first sight, positive $\alpha$ decreases both coordinate radii. However, the effective angular circumference is not $2\pi r$, but $2\pi\gamma(r)$. Therefore, the physical size of the shadow and the capture scale are not determined solely by the horizon's coordinate location.

The horizon circumference function is
\begin{equation}
 \gamma_h = \gamma(r_h) = \frac{1}{\sqrt{1+\alpha}}
 \left[ A^2+2\alpha\sqrt{1+\alpha}\,A(A+B)
 \right]^{1/2}.
 \label{eq:gamma_horizon}
\end{equation}
Thus, the Lorentz-violating correction modifies the acoustic geometry both radially and angularly. This combined effect—a smaller coordinate horizon but a non-trivially modified angular circumference—is the primary reason why the shadow size is not a monotone function of $\alpha$ alone.

\section{Null rays and shadow interval}
\label{sec:nullrays}

The acoustic shadow is determined by null rays satisfying
\begin{equation}
 g_{\mu\nu}\dot{x}^{\mu}\dot{x}^{\nu}=0 ,
 \label{eq:null_condition}
\end{equation}
where the dot denotes differentiation with respect to an affine parameter.

Because the metric is stationary and axisymmetric, there are two conserved quantities. We define
\begin{equation}
 E
 =
 {\cal H}(r)\dot{t}+B\dot{\phi},
 \qquad
 L
 =
 \gamma^2(r)\dot{\phi}-B\dot{t}.
 \label{eq:constants}
\end{equation}

The impact parameter is
\begin{equation}
 b=\frac{L}{E}.
 \label{eq:impact_parameter}
\end{equation}
Solving Eq.~\eqref{eq:constants} for $\dot{t}$ and $\dot{\phi}$ gives
\begin{equation}
 \frac{\dot{t}}{E}
 =
 \frac{\gamma^2(r)-Bb}{{\cal D}(r)},
 \qquad
 \frac{\dot{\phi}}{E}
 =
 \frac{B+{\cal H}(r)b}{{\cal D}(r)},
 \label{eq:tdot_phidot}
\end{equation}
where
\begin{equation}
 {\cal D}(r)
 =
 {\cal H}(r)\gamma^2(r)+B^2 .
 \label{eq:D_function}
\end{equation}
Substitution into the null condition gives the radial equation
\begin{equation}
 \frac{\dot{r}^2}{E^2}
 =
 \frac{{\cal G}(r)}{{\cal D}(r)}
 {\cal R}(r;b),
 \label{eq:radial_equation}
\end{equation}
with
\begin{equation}
 {\cal R}(r;b)
 =
 \gamma^2(r)-2Bb-{\cal H}(r)b^2 .
 \label{eq:R_function}
\end{equation}

The allowed region of motion is determined by ${\cal R}(r;b)\geq0$, with the prefactor ${\cal G}/{\cal D}$ fixing the radial normalization.

\begin{figure}[t]
\centering
\includegraphics[width=0.98\linewidth]{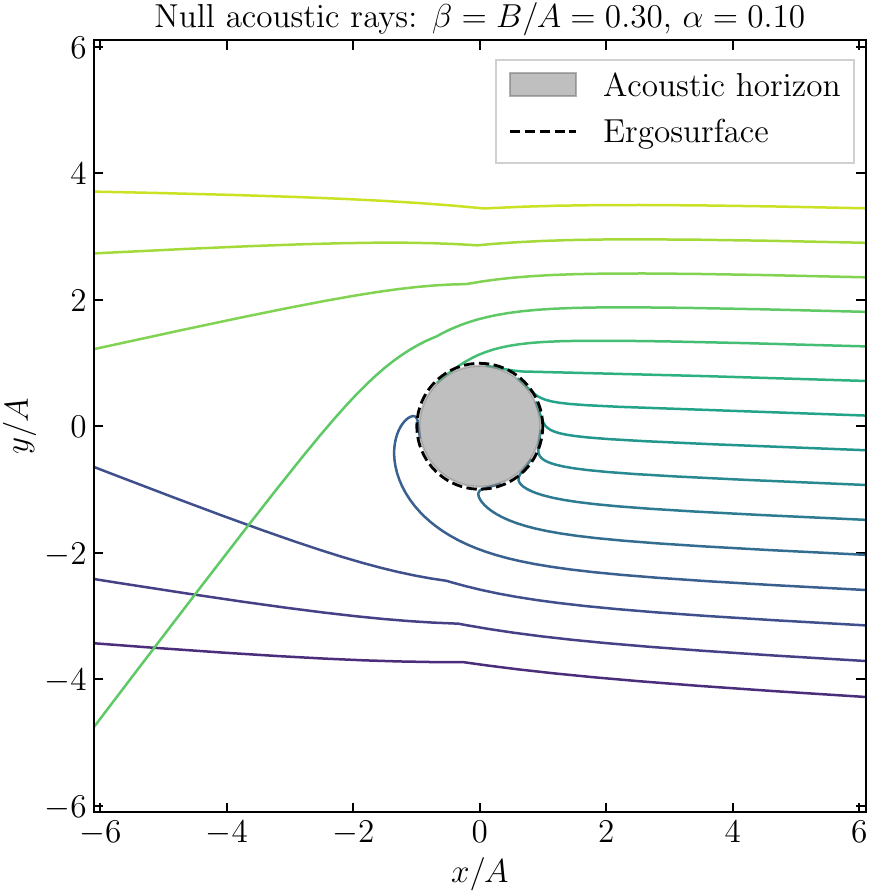}
\caption{Null acoustic rays for the Lorentz-violating rotating acoustic black hole with $\alpha=0.10$ and $B/A=0.30$.  The filled disk denotes the acoustic horizon $r_h=A/\sqrt{1+\alpha}$, while the dashed circle denotes the ergosurface.  The plot displays the same geometric role as the null-ray skeleton in the ordinary rotating draining-bathtub case, but with the Lorentz-deformed horizon and circumference function.}
\label{fig:lv_null_rays}
\end{figure}

The angular trajectory follows from
\begin{equation}
 \frac{d\phi}{dr}
 =
 \frac{B+{\cal H}(r)b}
 {\pm\sqrt{{\cal G}(r){\cal D}(r){\cal R}(r;b)}} .
 \label{eq:dphidr}
\end{equation}
The minus sign corresponds to an inward branch, and the plus sign corresponds to an outward branch after a turning point. A turning point satisfies ${\cal R}(r_{\rm tp};b)=0$, with $r_{\rm tp}>r_h $. Figure~\ref{fig:lv_null_rays} illustrates this ray classification for a representative Lorentz-violating rotating configuration.  The central filled disk represents the acoustic horizon, and the dashed circle denotes the ergosurface.  Rays launched with impact parameters inside the capture interval fall through the horizon, while rays outside the interval reach a turning point and are scattered.  Near the critical values, the trajectories wind around the acoustic vortex before escaping or being absorbed.

\subsection{Critical rays}

The boundary of the shadow is formed by unstable circular null rays. These satisfy the following conditions
\begin{equation}
 {\cal R}(r_c;b_c)=0,
 \qquad
 \left.
 \frac{\partial {\cal R}}{\partial r}
 \right|_{r=r_c,b=b_c}=0 .
 \label{eq:critical_conditions}
\end{equation}
Since we have
\begin{equation}
\frac{\partial{\cal R}}{\partial r}
= 2r+2\alpha(A+B)-{\cal H}'(r)b^2 ,
\label{eq:R_derivative}
\end{equation}
with ${\cal H}'(r)=\frac{2(A^2+B^2)}{r^3}$, the second critical condition gives
\begin{equation}
 b_c^2
 =
 \frac{\left[r_c+\alpha(A+B)\right]r_c^3}{A^2+B^2}.
 \label{eq:bc_implicit}
\end{equation}
Equations~\eqref{eq:critical_conditions} and \eqref{eq:bc_implicit} provide an implicit definition of the critical impact parameters. This form is robust and can be solved numerically without assuming that either $B/A$ or $\alpha$ is small. In particular, Eq.~\eqref{eq:bc_implicit} shows that for $\alpha>0$ the effective numerator $[r_c+\alpha(A+B)]r_c^3$ is larger than the $\alpha=0$ value $r_c^4$, so the critical impact parameter grows with the Lorentz-breaking correction, consistent with the broader shadow interval observed numerically.

In the notation commonly used in the absorption analysis, one may introduce a shifted impact parameter
\begin{equation}
 \mathfrak{b}=b+B.
 \label{eq:shifted_b}
\end{equation}
For small Lorentz violation, the critical values may be expanded as
\begin{align}
 \mathfrak{b}^{\pm}_{c}
 &\simeq-B \pm
 \frac{2\sqrt{A^2+B^2}}{\sqrt{1+\alpha}}
 \nonumber\\&-\frac{\alpha(A+B)\sqrt{2A^2+2B\left[B\mp\sqrt{A^2+B^2}\right]
 }}{B\mp\sqrt{A^2+B^2}} .
 \label{eq:bc_small_alpha}
\end{align}
This expansion is obtained by substituting $r_c = r_c^{(0)} + \alpha r_c^{(1)} + \mathcal{O}(\alpha^2)$ into Eqs.~\eqref{eq:critical_conditions} and \eqref{eq:bc_implicit}, where $r_c^{(0)} = \sqrt{2(A^2+B^2)}$ is the unperturbed critical radius, and collecting terms order by order. The argument of the square root in the correction term of Eq.~\eqref{eq:bc_small_alpha} is proportional to $2A^2 + 2B[B \mp \sqrt{A^2+B^2}]$, which is positive for both branches when $A\neq 0$: for the upper sign one has $B - \sqrt{A^2+B^2} < 0$, so the argument reduces to $2A^2 - 2B(\sqrt{A^2+B^2}-B) = 2(A^2+B^2) - 2B\sqrt{A^2+B^2} > 0$ since $\sqrt{A^2+B^2} > B$; the lower sign gives $2A^2 + 2B(B + \sqrt{A^2+B^2}) > 0$ trivially. The expansion is therefore real and well defined for both branches in the regime $\alpha\geq 0$, $A>0$.

For the finite-distance transfer profiles below, we use the asymptotically normalized screen coordinate measured by a static observer at infinity,
\begin{equation}
 X=\sqrt{1+\alpha}\,b .
 \label{eq:X_normalized}
\end{equation}
The factor $\sqrt{1+\alpha}$ arises naturally from the asymptotic structure of the metric. At large $r$, the effective acoustic line element approaches
\begin{equation}
 ds^2 \simeq (1+\alpha)\,dt^2 - dr^2 - r^2 d\phi^2,
 \label{eq:asymptotic_metric}
\end{equation}
so the sound speed measured at infinity is $c_\infty = 1/\sqrt{1+\alpha}$ relative to the coordinate time $t$. A static observer at infinity measures frequencies rescaled by $\sqrt{1+\alpha}$ relative to the coordinate energy $E$; see Eq.~\eqref{eq:observer_frequency}. The screen coordinate $X = \sqrt{1+\alpha}\,b$ is therefore the impact parameter as inferred from the asymptotically measured ray deflection angle, and it is the natural quantity to compare with what a physical detector registers. In the limit $\alpha\to0$, $X\to b$ recovers the standard screen coordinate of the undeformed draining-bathtub geometry.

Thus, the corresponding screen critical points are
\begin{equation}
 X_c^\pm\simeq \sqrt{1+\alpha}\,b_c^\pm
 =\sqrt{1+\alpha}\,(\mathfrak{b}_c^\pm-B) .
 \label{eq:Xcritical}
\end{equation}

\subsection{Centroid and width}

The acoustic shadow interval is
\begin{equation}
 X_c^- \leq X \leq X_c^+ .
 \label{eq:shadow_interval}
\end{equation}

We define the centroid and width by
\begin{equation}
 X_{\rm mid}^{(\alpha)}
 =
 \frac{X_c^+ + X_c^-}{2},
 \qquad
 \Delta X_\alpha
 =
 X_c^+-X_c^- .
 \label{eq:centroid_width}
\end{equation}

For $B=0$, the centroid vanishes, and the width becomes, to first order in $\alpha$,
\begin{equation}
 \Delta X_\alpha
 \simeq
 4A+2\alpha A(\sqrt{2}-1)+{\cal O}(\alpha^2).
 \label{eq:static_width}
\end{equation}

The coefficient $(\sqrt{2}-1)\approx 0.414$ follows from the combined expansion of the critical radius and of the asymptotic screen normalization. For $B=0$, the critical radius behaves as
\begin{equation}
 r_c
 =
 A\sqrt{2}
 +
 \alpha A\left(\frac{1}{4}-\frac{\sqrt{2}}{2}\right)
 +{\cal O}(\alpha^2),
\end{equation}
while the critical screen coordinate satisfies
\begin{equation}
 X_c^\pm
 =
 \pm
 \left[
 2A+\alpha A(\sqrt{2}-1)
 \right]
 +{\cal O}(\alpha^2).
\end{equation}
Therefore, Eq.~\eqref{eq:static_width} follows directly, and positive Lorentz violation broadens the static acoustic shadow at leading order.

For weak rotation, $\beta=B/A\ll1$, one obtains
\begin{equation}
 \frac{X_{\rm mid}^{(\alpha)}}{A}
 \simeq
 -2\beta+\left(1+\frac{1}{\sqrt{2}}\right)\alpha\beta
 +{\cal O}(\beta^2,\alpha^2),
 \label{eq:weak_centroid}
\end{equation}
and
\begin{equation}
 \frac{\Delta X_\alpha}{A}
 \simeq
 4
 +2\alpha(\sqrt{2}-1)
 +2\sqrt{2}\alpha\beta
 +{\cal O}(\beta^2,\alpha^2).
 \label{eq:weak_width}
\end{equation}
These expressions show that the width and centroid respond differently. The width is already sensitive to $\alpha$ when $B=0$, whereas the leading centroid shift is rotational and receives a mixed Lorentz-violating correction proportional to $\alpha B$.  The coefficient of the mixed term in Eq.~\eqref{eq:weak_centroid} is obtained by expanding the two critical branches simultaneously; this provides a useful check on the numerical root finder used below.

\begin{figure}[t]
\centering
\includegraphics[width=0.98\linewidth]{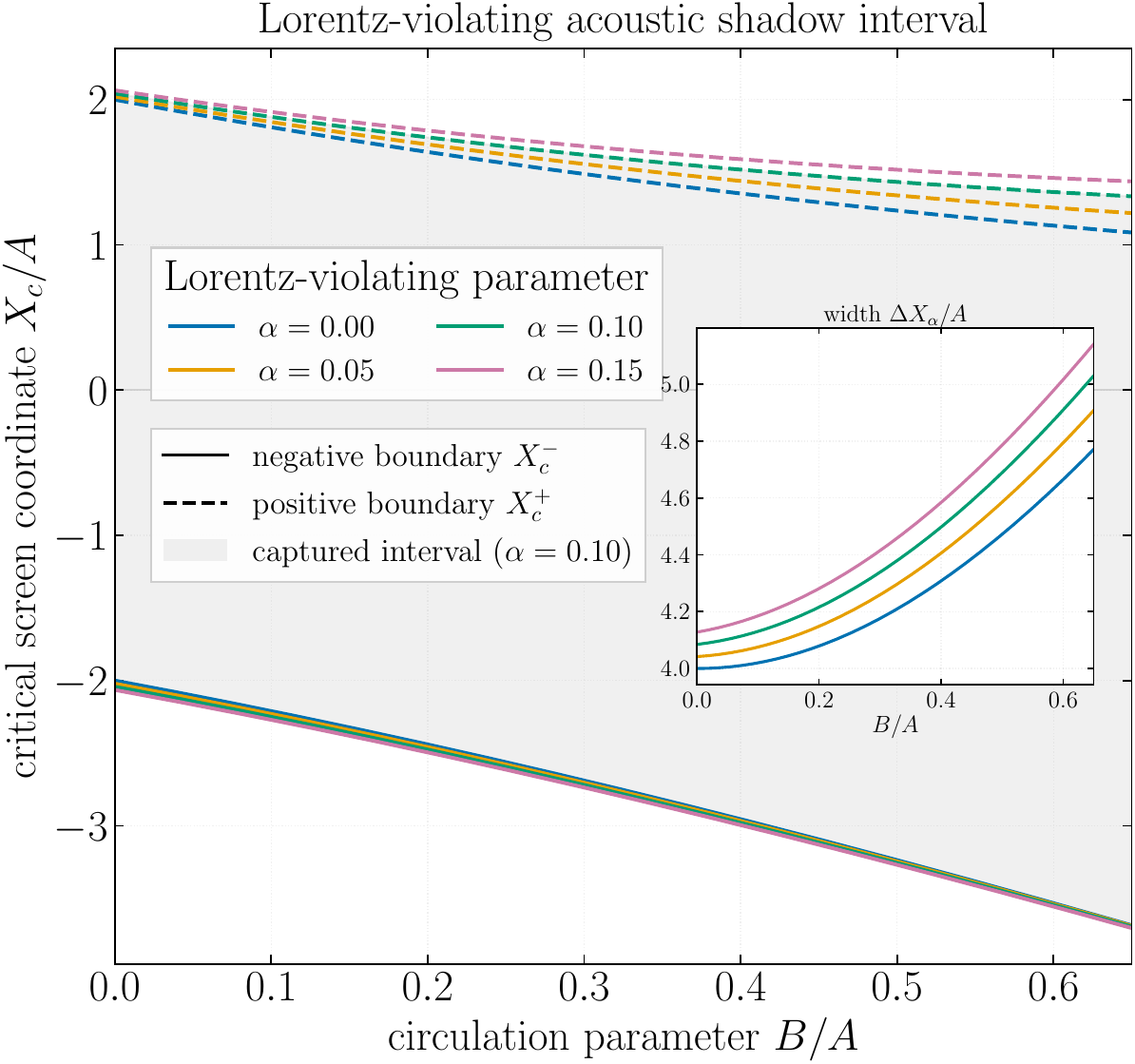}
\caption{Lorentz-violating acoustic shadow interval as a function of the circulation parameter.  The solid curves represent the negative critical boundary $X_c^-/A$, and the dashed curves represent the positive critical boundary $X_c^+/A$ for several values of $\alpha$.  The shaded band marks the captured interval for the representative case $\alpha=0.10$.  The inset shows the corresponding width $\Delta X_\alpha/A$, making the Lorentz-violation-induced broadening explicit.  Increasing $\alpha$ enlarges the global capture scale, whereas increasing $B/A$ mainly displaces the interval and enhances the branch asymmetry.}
\label{fig:lv_shadow_interval}
\end{figure}

The complete numerical critical interval is shown in Fig.~\ref{fig:lv_shadow_interval}.  The solid curves give the negative critical branch, and the dashed curves give the positive critical branch for several values of $\alpha$.  The shaded region highlights the capture interval for the representative case $\alpha=0.10$.  The figure makes clear that the Lorentz-breaking parameter changes the global width of the shadow, while increasing $B/A$ displaces the interval and makes the two branches asymmetric.

\section{Acoustic redshift}
\label{sec:redshift}

The acoustic frequency measured by an observer with four-velocity $u^\mu$ is
\begin{equation}
 \omega_{\rm meas}=k_\mu u^\mu ,
 \label{eq:measured_frequency}
\end{equation}
where $k^\mu=\dot{x}^\mu$ is the null tangent. The acoustic redshift factor is
\begin{equation}
 g_{\rm ac}
 =
 \frac{\omega_{\rm obs}}{\omega_{\rm em}}
 =
 \frac{(k_\mu u^\mu)_{\rm obs}}{(k_\mu u^\mu)_{\rm em}} .
 \label{eq:redshift_definition}
\end{equation}

A static observer at a large radius has
\begin{equation}
 u^\mu_{\rm obs}
 =
 \left(\frac{1}{\sqrt{1+\alpha}},0,0\right),
 \label{eq:observer_velocity}
\end{equation}
because ${\cal H}(r\to\infty)=1+\alpha$. The normalization condition $g_{\mu\nu}u^\mu_{\rm obs}u^\nu_{\rm obs}={\cal H}(r\to\infty)(u^t_{\rm obs})^2 = 1$ then fixes $u^t_{\rm obs}=1/\sqrt{1+\alpha}$, confirming Eq.~\eqref{eq:observer_velocity}. Therefore,
\begin{equation}
 (k_\mu u^\mu)_{\rm obs}
 =
 k_t u^t_{\rm obs}
 =
 \frac{E}{\sqrt{1+\alpha}} .
 \label{eq:observer_frequency}
\end{equation}

For the emitter, we use the general velocity field
\begin{equation}
 u^\mu_{\rm em}
 =
 u^t_{\rm em}
 \left(1,v^r_{\rm em},\Omega_{\rm em}\right),
 \label{eq:emitter_velocity}
\end{equation}
where $\Omega_{\rm em}=d\phi/dt$ is the emitter angular velocity and $v^r_{\rm em}=dr/dt$ is a possible radial coordinate velocity. The normalization condition gives
\begin{equation}
 \left(u^t_{\rm em}\right)^{-2}
 =
 {\cal H}(r)
 +2B\Omega_{\rm em}
 -\gamma^2(r)\Omega_{\rm em}^2
 -\frac{\left(v^r_{\rm em}\right)^2}{{\cal G}(r)} .
 \label{eq:ut_normalization}
\end{equation}

Using $k_t=E$ and $k_\phi=-L=-bE$, we obtain
\begin{equation}
 g_{\rm ac}
 =
 \frac{
 1
 }
 {
 \sqrt{1+\alpha}\,
 u^t_{\rm em}
 \left[
 1-b\Omega_{\rm em}
 +(k_r/E)v^r_{\rm em}
 \right]
 } .
 \label{eq:gac_general}
\end{equation}

The radial covariant momentum is branch-dependent. Since $k_r=g_{rr}\dot r=-\dot r/{\cal G}(r)$, one obtains
\begin{equation}
 \frac{k_r}{E}
 =
 -\,\mathrm{sgn}(\dot r)
 \sqrt{
 \frac{{\cal R}(r;b)}
 {{\cal G}(r){\cal D}(r)}
 } .
 \label{eq:kr_branch}
\end{equation}
Equivalently,
\begin{equation}
 \frac{k_r}{E}
 =
 \begin{cases}
 +\sqrt{\dfrac{{\cal R}(r;b)}
 {{\cal G}(r){\cal D}(r)}} ,
 & \dot r<0 \quad \text{inward branch},\\[1.2em]
 -\sqrt{\dfrac{{\cal R}(r;b)}
 {{\cal G}(r){\cal D}(r)}} ,
 & \dot r>0 \quad \text{outward branch}.
 \end{cases}
\end{equation}
This sign change is the mechanism by which left-right branch asymmetry enters the redshift observable. For circular emitters, $v^r_{\rm em}=0$, the radial branch drops out and the redshift simplifies to
\begin{equation}
 g_{\rm ac}
 =
 \frac{
 \sqrt{
 {\cal H}(r)
 +2B\Omega_{\rm em}
 -\gamma^2(r)\Omega_{\rm em}^2
 }
 }
 {
 \sqrt{1+\alpha}
 \left(1-b\Omega_{\rm em}\right)
 } .
 \label{eq:gac_circular}
\end{equation}

The Lorentz-violating parameter affects this redshift in several coupled ways: it changes the asymptotic normalization through the factor $\sqrt{1+\alpha}$ in the denominator, modifies the effective gravitational term ${\cal H}(r)$ in the numerator, and changes the angular circumference function $\gamma(r)$ that controls the centrifugal term. These three modifications are not degenerate with each other, so the same physical source profile can produce distinguishably different observed frequency maps for different values of $\alpha$ even at fixed $B$.

Useful emitter models include
\begin{equation}
 \Omega_{\rm em}(r)=0,
 \qquad
 \Omega_{\rm em}(r)=\Omega_0 r^{-q},
 \qquad
 \Omega_{\rm em}(r)=\frac{B}{\gamma^2(r)} .
 \label{eq:omega_models}
\end{equation}
The first describes static emitters, the second a phenomenological rotating disk, and the third a vortex-comoving angular profile in which the emitter rotates with the local frame-dragging angular velocity.

\section{Intensity transfer}
\label{sec:transfer}

The observed intensity is written as a transfer relation rather than as a fixed source model. For a thin emitting ring, each ray can intersect the source on one or more branches. The observed intensity is
\begin{equation}
 I_{\rm obs}(X)
 =
 \sum_n
 g_{{\rm ac},n}^{\eta}
 I_{\rm em}(r_n,\phi_n)
 J_n^{-1},
 \label{eq:thin_transfer}
\end{equation}
where $n$ labels the intersections, $\eta$ is the transfer exponent, and $J_n^{-1}$ is the local source-to-screen magnification.

The exponent $\eta$ is kept explicit. For electromagnetic specific intensity in four-dimensional relativistic transfer, the familiar invariant scaling is $I_\nu/\nu^3={\rm constant}$~\cite{Lindquist1966,Chandrasekhar1983}, suggesting $\eta=3$. For an ideal two-dimensional acoustic energy-flux measurement, a scaling closer to $\eta=2$ may be more natural, because in 2+1 dimensions the phase-space measure differs from the four-dimensional case. In a laboratory, however, the effective exponent depends on the detector response and reconstruction protocol. Therefore, $\eta$ should be treated as part of the transfer convention and reported explicitly when quoting asymmetry observables.

The intrinsic emissivity is written as
\begin{equation}
 I_{\rm em}(r,\phi)
 =
 I_0 R(r)P(\phi).
 \label{eq:emissivity}
\end{equation}

For an extended disk, we take
\begin{equation}
 R(r)
 =
 \left(\frac{r}{r_h}\right)^{-p}
 \Theta(r-r_{\rm in})
 \Theta(r_{\rm out}-r).
 \label{eq:radial_emissivity}
\end{equation}

A useful scaling is
\begin{equation}
 r_{\rm in}=\chi_{\rm in}r_h,
 \qquad
 r_{\rm out}=\chi_{\rm out}r_h,
 \label{eq:disk_scaling}
\end{equation}

so that the emitting region follows the Lorentz-violating horizon scale. This choice ensures that a variation in $\alpha$ rescales the emitting region along with the horizon, keeping the geometric ratio between the source and capture radii approximately fixed.

Nonaxisymmetric emission can be modeled by
\begin{equation}
 P(\phi)
 =
 1+\sum_{m=1}^{m_{\rm max}}
 \epsilon_m
 \cos\left[m(\phi-\phi_m)\right],
 \qquad
 |\epsilon_m|<1 .
 \label{eq:angular_emissivity}
\end{equation}
This term is important because an intrinsic hot spot can imitate a brightness asymmetry. Thus, brightness asymmetry alone should not be identified with rotation or Lorentz violation; a consistency check with the geometric and redshift observables is always required.

For a thin ring at $r=r_s$, the one-dimensional Jacobian is
\begin{equation}
 J^{-1}_{\rm ring}(b)
 =
 \gamma(r_s)
 \left|
 \frac{d\phi_{\rm em}}{db}
 \right|.
 \label{eq:ring_jacobian}
\end{equation}
The factor $\gamma(r_s)$ replaces the ordinary radius $r_s$ because the physical angular length is determined by the Lorentz-violating circumference function. Near caustics, $J^{-1}_{\rm ring}$ can become large. Therefore, a raw infinitesimal ring should be treated as a diagnostic limit, not as a final observable.

A more physical ring has finite radial width,
\begin{equation}
 R_{\rm ring}(r)
 =
 \exp\left[
 -\frac{(r-r_s)^2}{2\sigma_r^2}
 \right].
 \label{eq:finite_ring}
\end{equation}
For an extended optically thin disk, the observed intensity is obtained by direct integration along the ray:
\begin{equation}
 I_{\rm obs}(X)
 =
 \sum_{\rm branches}
 \int_{\lambda_{\rm in}}^{\lambda_{\rm out}}
 g_{\rm ac}^{\eta}
 I_{\rm em}\!\left[r(\lambda),\phi(\lambda)\right]
 W\!\left[r(\lambda)\right]
 d\lambda .
 \label{eq:disk_transfer}
\end{equation}
The window function $W(r)$ restricts the integration to the emitting region. Finally, finite detector resolution is modeled by convolution:
\begin{equation}
 I_{\rm obs}^{\rm det}(X)
 =
 \int dX'\,
 I_{\rm obs}(X')
 \frac{1}{\sqrt{2\pi}\sigma_X}
 \exp\left[
 -\frac{(X-X')^2}{2\sigma_X^2}
 \right].
 \label{eq:detector_convolution}
\end{equation}

\begin{figure}[t]
\centering
\includegraphics[width=0.98\linewidth]{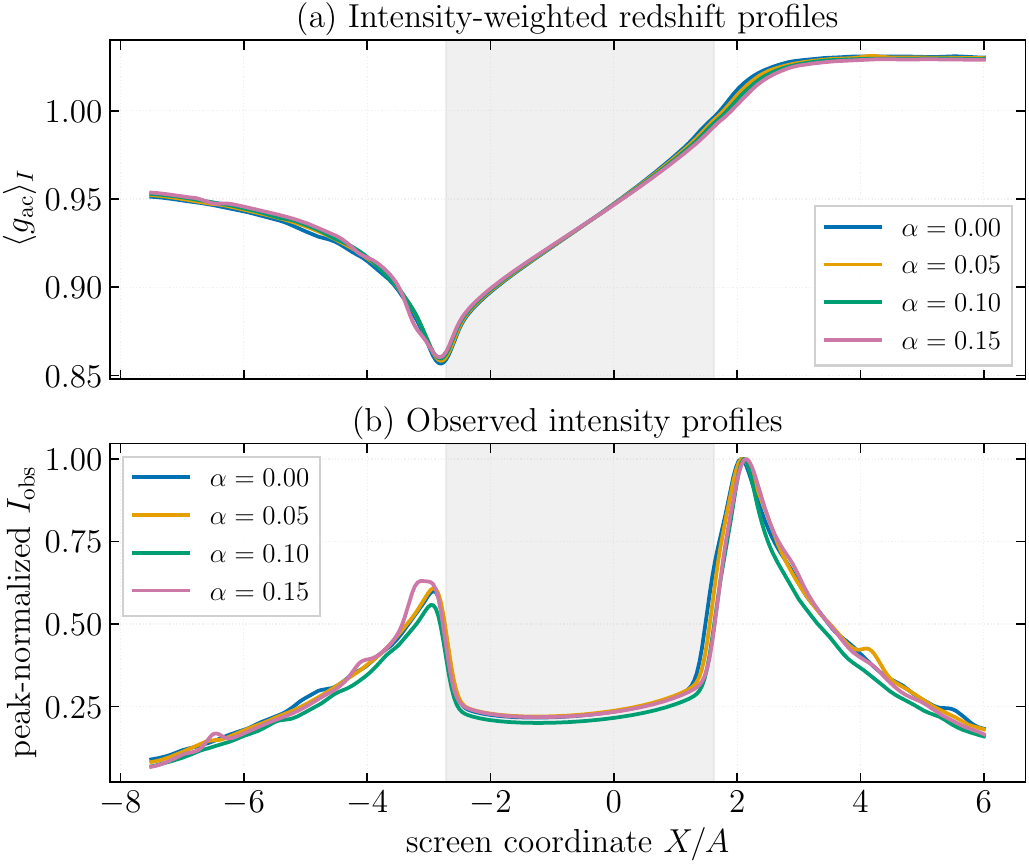}
\caption{Direct extended-disk transfer profiles for the Lorentz-violating rotating acoustic black hole.  The circulation is fixed at $B/A=0.30$, and the Lorentz-breaking parameter is varied.  The upper panel shows the intensity-weighted acoustic redshift $\langle g_{\rm ac}\rangle_I$, while the lower panel shows the observed intensity normalized by the peak of each curve.  The shaded region denotes the representative capture interval for $\alpha=0.10$.}
\label{fig:lv_transfer_profiles}
\end{figure}

The corresponding finite-resolution transfer profiles are displayed in Fig.~\ref{fig:lv_transfer_profiles}.  The upper panel shows the intensity-weighted acoustic redshift and the lower panel shows the peak-normalized observed intensity for a fixed circulation $B/A=0.30$ and several Lorentz-violating parameters.  The gray band marks the representative geometric capture interval.  The redshift profile retains a left-right tilt because the background rotates, while the variation with $\alpha$ changes the global normalization and the detailed branch weights.  The intensity curves are normalized to their own maxima to compare morphology rather than absolute power.

\begin{figure}[tbhp]
\centering
\includegraphics[width=0.98\linewidth]{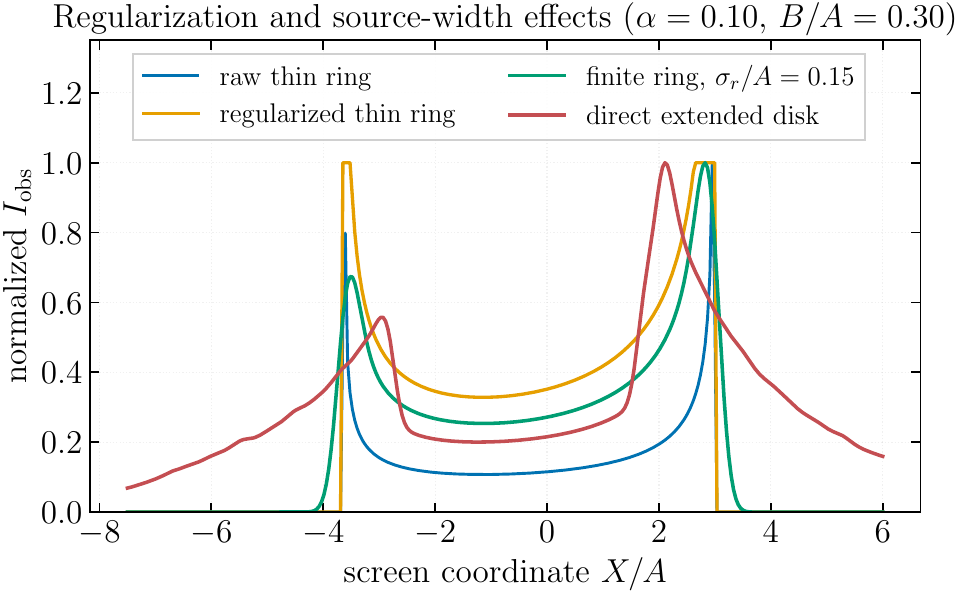}
\caption{Regularization and source-width effects for $\alpha=0.10$ and $B/A=0.30$.  The raw thin-ring profile is dominated by caustic magnification.  The regularized ring and finite-width ring suppress the singular peaks, whereas the direct extended-disk curve yields the finite-resolution profile used in the main observable discussion.}
\label{fig:lv_regularization}
\end{figure}

Figure~\ref{fig:lv_regularization} compares the raw thin-ring limit with regularized and finite-source profiles.  The infinitesimal ring contains sharp caustic enhancements generated by the one-dimensional source-to-screen Jacobian.  These features are useful for diagnosing the map, but they are not stable finite-resolution observables.  Capping the Jacobian, adding a finite radial width, or integrating through an extended disk smooths the singular structure.  The direct disk profile is therefore the appropriate baseline for the transfer diagnostics.

\section{Two-dimensional acoustic screen maps}
\label{sec:2dscreen}

The 1D profiles $I_{\rm obs}(X)$ of the preceding sections contain the actual ray-tracing information of the $(2+1)$-dimensional draining-bathtub geometry. For visualization purposes, we now construct a synthetic two-dimensional screen by supplementing the physical impact-parameter coordinate with a transverse coordinate $b_y$. This does not introduce an independent geodesic degree of freedom in the strict $(2+1)$D geometry; instead, it provides a phenomenological way to display the one-dimensional capture interval as a strip and to model a finite transverse source profile.

\subsection{Shadow geometry in the two-dimensional screen plane}

In the (2+1)D acoustic analog, the fluid and all null rays lie in a single spatial plane. A distant observer at $(r_{\rm obs}, \phi_{\rm obs})$ with $r_{\rm obs}\to\infty$ carries a one-dimensional screen parameterized by the impact parameter $b$ (equivalently $X = \sqrt{1+\alpha}\,b$). The screen direction is the unit vector perpendicular to the line of sight in the fluid plane.

To construct a two-dimensional image, one introduces a synthetic transverse coordinate $b_y$ alongside $b_x = b = X/\sqrt{1+\alpha}$. The physical ray tracing is still controlled by the longitudinal impact parameter $b_x$, while $b_y$ only parameterizes the phenomenological transverse source profile. The two-dimensional observed intensity is therefore
\begin{equation}
 I_{\rm obs}(b_x,b_y)
 =
 I_{\rm obs}^{\rm 1D}(b_x)\,W_y(b_y),
 \label{eq:2d_intensity}
\end{equation}
where $W_y(b_y)$ is a window encoding the finite extent of the acoustic source in the transverse direction. A natural choice motivated by the power-law radial profile is
\begin{equation}
 W_y(b_y)
 =
 \left[1+\frac{b_y^2}{b_x^2+\sigma_y^2}\right]^{-p/2},
 \label{eq:Wy}
\end{equation}
which ensures that emission is concentrated near the acoustic plane and falls off away from the equatorial direction.

The shadow in the two-dimensional $(b_x,b_y)$ screen plane is a \emph{vertical strip},
\begin{align}
 \text{Shadow}_{2D}
 =
 \Bigg\{
 (b_x,b_y)\;&:\;
 \frac{X_c^-}{\sqrt{1+\alpha}}
 \leq
 b_x \notag \\&
 \leq
 \frac{X_c^+}{\sqrt{1+\alpha}},
 \quad
 b_y\in\mathbb{R}
 \Bigg\}.
 \label{eq:shadow_2d}
\end{align}
This strip geometry is the natural synthetic two-dimensional representation of the $(2+1)$D capture interval. The strip width $\Delta X_\alpha/\sqrt{1+\alpha}$ is the Lorentz-violation-sensitive broadening discussed in Section~\ref{sec:nullrays}, and the strip centroid $X_{\rm mid}^{(\alpha)}/\sqrt{1+\alpha}$ is the rotation-sensitive displacement. For $B=0$ the strip is centered at the origin; for $B\neq0$ it is displaced toward negative $b_x$ by the amount $|X_{\rm mid}^{(\alpha)}|/\sqrt{1+\alpha}$.

\begin{figure*}[tbhp]
\centering
\includegraphics[width=0.82\textwidth]{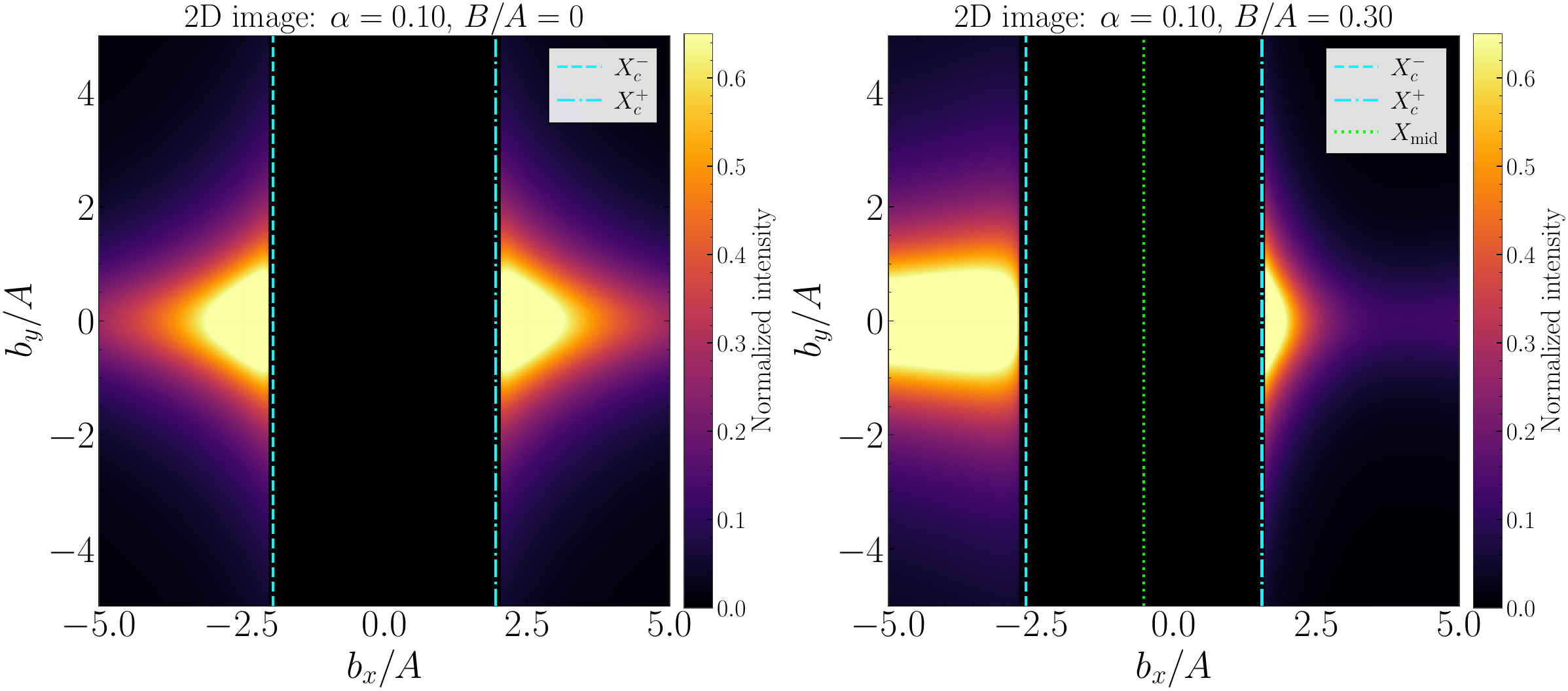}
\caption{Two-dimensional acoustic screen images in the $(b_x,b_y)$ plane for $\alpha=0.10$.
\emph{Left:} Nonrotating case, $B/A=0$, where the acoustic shadow is a symmetric dark strip of width $\Delta X_\alpha/\sqrt{1+\alpha}$ flanked by two brightness lobes.
\emph{Right:} Rotating case, $B/A=0.30$, where the shadow strip is displaced toward negative $b_x$ by the centroid shift $X_{\rm mid}^{(\alpha)}/\sqrt{1+\alpha}$, marked by the green dotted line. The two brightness lobes become asymmetric, with the approaching side brighter than the receding side, consistent with the branch-dependent flux imbalance found in the 1D transfer profiles. Cyan dashed lines indicate the critical boundaries $X_c^\pm/\sqrt{1+\alpha}$, and the intensity is normalized to the peak value in each panel.}
\label{fig:lv_2d_screen_map}
\end{figure*}

Figure~\ref{fig:lv_2d_screen_map} illustrates the two-dimensional screen images. The left panel shows the axisymmetric case ($B/A=0$, $\alpha=0.10$): the shadow appears as a symmetric dark strip flanked by bright emission lobes whose intensity decreases with increasing $|b_y|$. The right panel shows the rotating case ($B/A=0.30$, $\alpha=0.10$): the strip shifts toward negative $b_x$ (centroid displacement), and the two lobes become asymmetric in brightness, with the left lobe (approaching side) visibly brighter than the right lobe (receding side). The centroid line $b_x = X_{\rm mid}^{(\alpha)}/\sqrt{1+\alpha}$ (green dotted line) lies inside the dark strip, illustrating that the rotation-induced shift is a rigid displacement of the entire shadow interval.

\subsection{Observer-azimuth map for non-axisymmetric sources}

A second natural two-dimensional representation is obtained by sweeping the observer azimuthal angle $\phi_{\rm obs}\in[0,2\pi)$ while keeping the screen coordinate $X$. For a source with azimuthal modulation $P(\phi)=1+\epsilon\cos(\phi-\phi_0)$ (a localized hot spot), the emission angle at the source is
\begin{equation}
 \phi_{\rm em}(b,\phi_{\rm obs})
 =
 \phi_{\rm obs}-\Delta\phi(b),
 \label{eq:phi_em_2d}
\end{equation}
where the angular shift accumulated along the null ray from $r_{\rm obs}$ to the source ring at $r=r_s$ is
\begin{equation}
 \Delta\phi(b)
 =
 \int_{r_s}^{r_{\rm obs}}
 \frac{B+{\cal H}(r)b}
 {\sqrt{{\cal G}(r){\cal D}(r){\cal R}(r;b)}}
 \,dr .
 \label{eq:delta_phi}
\end{equation}

Since the metric is independent of $\phi$, $\Delta\phi(b)$ does not depend on $\phi_{\rm obs}$. The two-dimensional observed intensity is therefore
\begin{equation}
 I(X,\phi_{\rm obs})
 =
 \sum_n
 g_{{\rm ac},n}^\eta\,
 \bigl[1+\epsilon\cos(\phi_{\rm obs}-\Delta\phi_n-\phi_0)\bigr]\,
 J_n^{-1} ,
 \label{eq:I_2d_phi}
\end{equation}
where the sum runs over ray branches $n$ that reach the source ring. For $\epsilon=0$ this reduces to the 1D profile multiplied by a constant. For $\epsilon\neq0$, the map $I(X,\phi_{\rm obs})$ shows sinusoidal modulation in $\phi_{\rm obs}$ at fixed $X$, with a phase and amplitude that depend on $\Delta\phi(b)$.

\begin{figure*}[tbhp]
\centering
\includegraphics[width=0.92\textwidth]{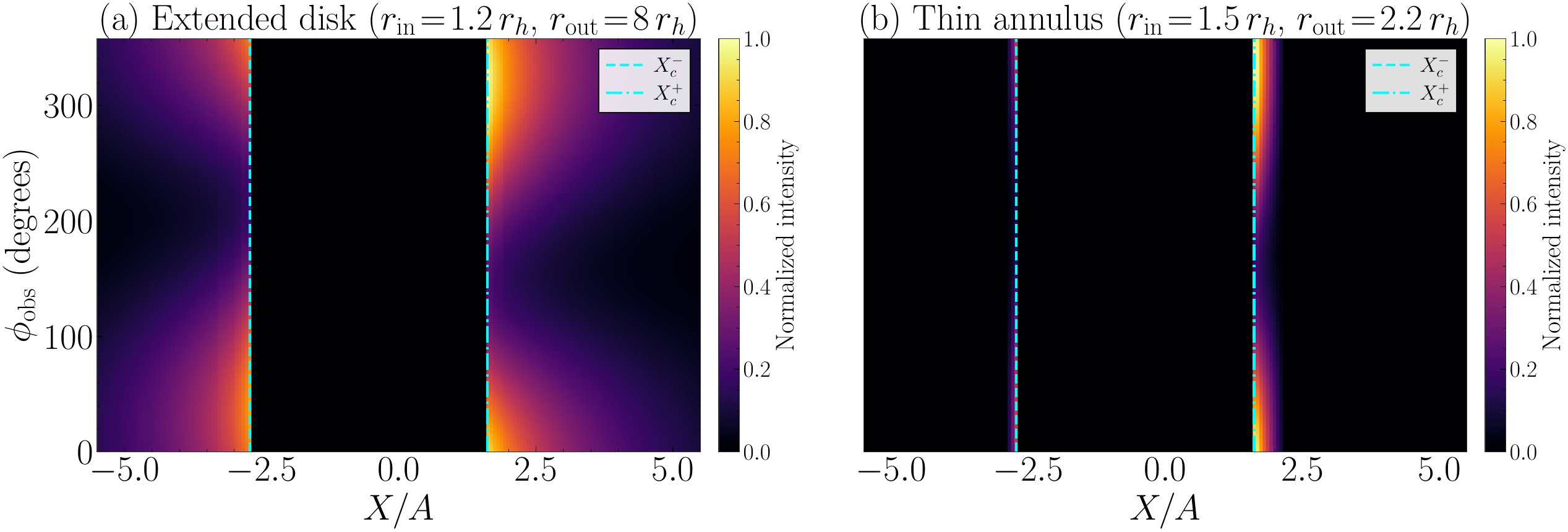}
\caption{Observer-azimuth intensity map $I(X,\phi_{\rm obs})$ for $\alpha=0.10$, $B/A=0.30$, and hot-spot amplitude $\epsilon=0.7$ with $\phi_0=0$. Panel (a) shows the extended disk ($r_{\rm in}=1.2r_h$, $r_{\rm out}=8r_h$): the shadow strip (dark band) is clearly visible, and the approaching lobe ($X<X_c^-$) is brighter than the receding lobe ($X>X_c^+$) due to the rotational flux asymmetry. The azimuthal hot-spot modulation is suppressed because the emission is averaged over many radii. Panel (b) shows the thin annulus ($r_{\rm in}=1.5r_h$, $r_{\rm out}=2.2r_h$): the hot-spot modulation is now clearly visible as a smooth sinusoidal gradient in $\phi_{\rm obs}$, with maximum brightness near $\phi_{\rm obs}=0$ (observer facing the hot spot) and minimum near $\phi_{\rm obs}=180^\circ$. The comparison between the two panels confirms that extended sources suppress the $\phi_{\rm obs}$-modulation, whereas concentrated rings amplify it. Cyan lines mark the critical boundaries $X_c^\pm$.}
\label{fig:lv_2d_obs_map}
\end{figure*}

Figure~\ref{fig:lv_2d_obs_map} shows this observer-azimuth map for $\alpha=0.10$, $B/A=0.30$ and hot-spot amplitude $\epsilon=0.7$. Panel (a) uses an extended disk, which suppresses the $\phi_{\rm obs}$-modulation through radial averaging but clearly displays the rotation-induced flux asymmetry between the two lobes. Panel (b) uses a thin annulus, where the hot-spot modulation is visible as a smooth sinusoidal gradient in $\phi_{\rm obs}$: the intensity peaks near $\phi_{\rm obs}=0$ (observer facing the hot spot) and is minimum near $\phi_{\rm obs}=180^\circ$. The shadow strip appears as a dark vertical band in both panels. This comparison illustrates a general principle: an extended source suppresses $\phi_{\rm obs}$-modulation compared to a thin ring, mirroring the regularization effect already seen in the 1D caustic analysis. An observer who sees the brightest image at a given $\phi_{\rm obs}$ is the one for whom the hot spot contributes most to the ray crossing the source region. This modulation is the acoustic analogue of the time-variable brightness seen in hot-spot observations of real black holes by very long baseline interferometry~\cite{EHT2019VI}.

\subsection{Structural difference from three-dimensional imaging}

It is important to note that the strip shadow of Eq.~\eqref{eq:shadow_2d} is a synthetic two-dimensional representation of a genuinely one-dimensional capture interval in the $(2+1)$D calculation. In a cylindrically extended acoustic system, one would still expect a strip-like structure in the corresponding screen plane if the additional direction decouples from the in-plane dynamics. A circular shadow disk, familiar from Schwarzschild and Kerr black holes, requires a spherically (or quasi-spherically) symmetric geometry. Extending the present framework to a truly spherical acoustic geometry would require a different underlying flow model and falls outside the scope of the present work. The strip geometry displayed here should therefore be understood as the natural two-dimensional visualization of the $(2+1)$D rotating acoustic black-hole capture interval with Lorentz symmetry violation.

\section{Differential observables}
\label{sec:observables}

The main observables are divided into geometric, redshift and flux diagnostics.

\subsection{Geometric observables}

The geometric observables are the shadow centroid and width:
\begin{equation}
 X_{\rm mid}^{(\alpha)}
 =
 \frac{X_c^+ + X_c^-}{2},
 \qquad
 \Delta X_\alpha
 =
 X_c^+-X_c^- .
 \label{eq:geometric_observables}
\end{equation}

It is also useful to define the Lorentz-induced broadening relative to the ordinary draining-bathtub value:
\begin{equation}
 {\cal W}_{\alpha}
 =
 \frac{\Delta X_\alpha-\Delta X_0}{\Delta X_0},
 \label{eq:width_broadening}
\end{equation}
where $\Delta X_0\equiv \Delta X_{\alpha=0}(B)$ is evaluated at the same circulation parameter.

Similarly, the Lorentz correction to the centroid can be written as
\begin{equation}
 {\cal C}_{\alpha}
 =
 \frac{X_{\rm mid}^{(\alpha)}-X_{\rm mid}^{(0)}}{A}.
 \label{eq:centroid_correction}
\end{equation}

The quantity ${\cal W}_{\alpha}$ can be nonzero even for $B=0$, while ${\cal C}_{\alpha}$ is primarily a mixed Lorentz-rotation correction that vanishes when either $\alpha=0$ or $B=0$.

\subsection{Flux observables}

The branch-integrated fluxes are
\begin{equation}
 F_{\pm}
 =
 \int_{\pm(X-X_0)>0}
 I_{\rm obs}^{\rm det}(X)\,dX .
 \label{eq:branch_fluxes}
\end{equation}

The split point $X_0$ can be chosen in two ways:
\begin{equation}
 X_0 =
 \begin{cases}
 0,
 & \text{laboratory split},\\[0.4em]
 X_{\rm mid}^{(\alpha)},
 & \text{centroid-corrected split}.
 \end{cases}
 \label{eq:split_choices}
\end{equation}

The laboratory split measures what a fixed detector sees. The centroid-corrected split removes the pure geometric translation before comparing the two sides, thereby isolating the branch-dependent brightness contrast from the geometric shift caused by $B$.

The flux asymmetry is
\begin{equation}
 A_I^{\rm flux}
 =
 \frac{|F_+-F_-|}{F_++F_-}.
 \label{eq:flux_asymmetry}
\end{equation}

The peak asymmetry is
\begin{equation}
 A_I^{\rm peak}
 =
 \frac{
 |I_{\rm max}^{(+)}-I_{\rm max}^{(-)}|
 }
 {
 I_{\rm max}^{(+)}+I_{\rm max}^{(-)}
 } .
 \label{eq:peak_asymmetry}
\end{equation}

The peak asymmetry is generally more sensitive to caustics and finite resolution than the integrated flux asymmetry.

\subsection{Redshift observables}

The intensity-weighted redshift averages on the two sides of the image are
\begin{equation}
 \bar{g}_{\pm}
 =
 \frac{
 \int_{\pm(X-X_0)>0}
 g_{\rm ac}(X)I_{\rm obs}^{\rm det}(X)dX
 }
 {
 \int_{\pm(X-X_0)>0}
 I_{\rm obs}^{\rm det}(X)dX
 } .
 \label{eq:gbar_pm}
\end{equation}

The left-right redshift asymmetry is then
\begin{equation}
 A_g^{\rm LR}
 =
 \frac{|\bar{g}_+-\bar{g}_-|}{\bar{g}_++\bar{g}_-}.
 \label{eq:redshift_asymmetry}
\end{equation}

For an axisymmetric source and $B=0$, $A_g^{\rm LR}=0$ by symmetry.

A separate diagnostic is the global redshift contrast,
\begin{equation}
 C_g
 =
 \frac{g_{90}-g_{10}}{g_{90}+g_{10}},
 \label{eq:global_contrast}
\end{equation}
where $g_{10}$ and $g_{90}$ are the intensity-weighted $10\%$ and $90\%$ quantiles of the redshift distribution. Unlike $A_g^{\rm LR}$, this quantity need not vanish for $B=0$ because an extended disk contains emission from different radii.

\section{Physical discussion}
\label{sec:discussion}

The Lorentz-violating rotating acoustic black hole contains two different physical deformations. The circulation parameter $B$ is responsible for rotation, frame dragging, the displacement of the critical interval, and the branch-dependent Doppler imbalance. The Lorentz-breaking parameter $\alpha$ modifies the effective acoustic geometry itself. It changes the horizon radius, the ergosurface radius, the circumference function, and the normalization of the acoustic frequency measured at infinity.

This means that the shadow width and shadow centroid should not be interpreted in the same way. The width is sensitive to the global capture scale and can increase even when the system is nonrotating. Therefore, an enlarged acoustic shadow is not automatically a pure rotation signal. By contrast, the centroid shift is primarily a rotation signal. However, the mixed term proportional to $\alpha B$ shows that the calibration of the centroid as a circulation estimator is affected by Lorentz violation.

The redshift map provides an intermediate diagnostic. A nonzero left-right redshift asymmetry is difficult to generate from $\alpha$ alone if the source is axisymmetric and nonrotating. It is mainly caused by the combination of circulation, emitter motion, and the branch structure of the ray map. Therefore, $A_g^{\rm LR}$ should be used to check whether the inferred circulation from the centroid is consistent with the transferred frequency shifts.

The flux profile is the most detector-dependent observable. It contains the redshift factor $g_{\rm ac}^{\eta}$, the intrinsic emissivity, the source-to-screen Jacobian, and the detector response. A large brightness asymmetry can be produced by rotation, but it can also be modified by intrinsic azimuthal emissivity. For this reason, $A_I^{\rm flux}$ is more robust than $A_I^{\rm peak}$, but neither should be used without the geometric and redshift diagnostics.

The safest inference strategy is therefore multi-observable:
\begin{equation}
 \left\{
 \Delta X_\alpha,\,
 X_{\rm mid}^{(\alpha)},\,
 A_g^{\rm LR},\,
 A_I^{\rm flux},\,
 A_I^{\rm peak},\,
 C_g
 \right\}.
 \label{eq:diagnostic_hierarchy}
\end{equation}

The width constrains the Lorentz-violating broadening. The centroid constrains the rotation. The redshift asymmetry tests the branch-dependent transfer. The flux asymmetry quantifies how strongly this transfer is reflected in the observed intensity. The peak asymmetry is useful but more fragile. The global redshift contrast measures the spread of acoustic redshifts across the source and should not be confused with a pure rotation signal.

\begin{figure}[tbhp]
\centering
\includegraphics[width=0.98\linewidth]{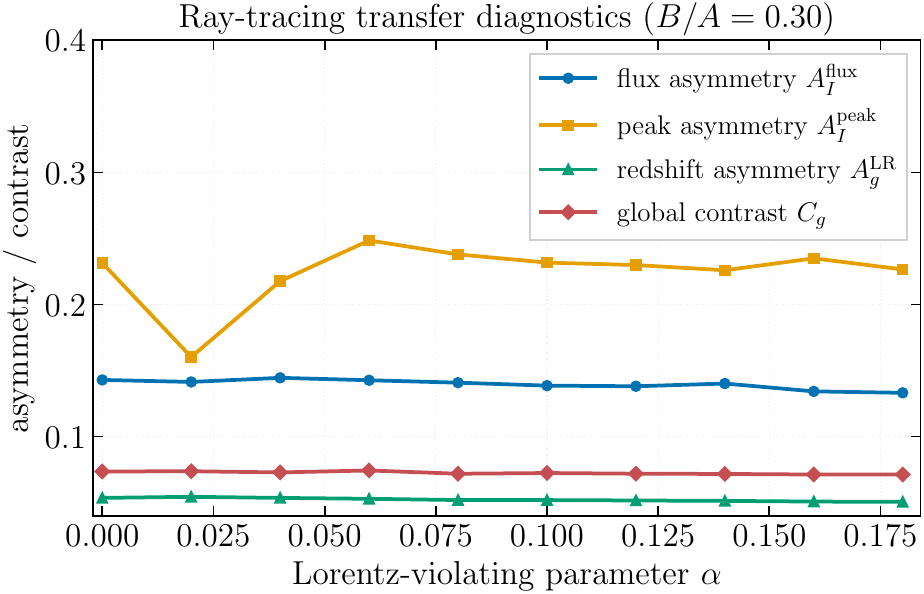}
\caption{Differential transfer observables as functions of the Lorentz-violating parameter at fixed $B/A=0.30$.  The plotted quantities are the branch-integrated flux asymmetry $A_I^{\rm flux}$, the peak asymmetry $A_I^{\rm peak}$, the left-right redshift asymmetry $A_g^{\rm LR}$, and the global redshift contrast $C_g$.  The flux and peak diagnostics are more transfer-dependent, while the redshift asymmetry directly probes the branch imbalance.}
\label{fig:lv_transfer_diagnostics}
\end{figure}

Figure~\ref{fig:lv_transfer_diagnostics} condenses the finite-resolution profiles into differential observables as functions of $\alpha$ at fixed $B/A=0.30$.  The flux and peak asymmetries depend on the transfer exponent and on the detailed source model, while the left-right redshift asymmetry is a cleaner branch diagnostic.  The global redshift contrast varies more weakly because it mainly measures the radial spread of redshifts across the extended emitting region.  These trends are consistent with the interpretation that $\alpha$ mostly recalibrates the capture and redshift scale, whereas $B$ is responsible for the dominant branch imbalance.

\subsection{Expected parameter trends}

For positive $\alpha$, the high-frequency capture scale is expected to grow, in agreement with the absorption analysis of Ref.~\cite{Campos2026}. In the image plane, this appears as a broader acoustic shadow interval. For increasing $B/A$, the interval is shifted to one side, and the two branches become more asymmetric. When both parameters are nonzero, mixed $\alpha B$ terms change the centroid calibration and alter the redshift and flux contrast between the two sides.

Table~\ref{tab:parameter_trends} summarizes the qualitative trends.
\begin{table*}[t]
\centering
\caption{Qualitative parameter trends for the Lorentz-violating transfer observables.}
\begin{tabular}{lccc}
\toprule
Parameter change & Shadow width & Shadow centroid & Transfer asymmetry \\
\midrule
Increase $\alpha$ at $B=0$ & increases & unchanged & no left-right tilt for symmetric source \\
Increase $B$ at $\alpha=0$ & increases weakly & shifts linearly & redshift and flux imbalance \\
Increase both $\alpha$ and $B$ & increases & shifted with correction & modified branch contrast \\
Change $\eta$ & unchanged & unchanged & changes intensity contrast \\
Add $P(\phi)\neq1$ & unchanged & unchanged & can mimic flux asymmetry \\
\bottomrule
\end{tabular}
\label{tab:parameter_trends}
\end{table*}

Thus, the geometry should be fitted before interpreting the brightness contrast.

\subsection{Numerical implementation}

A practical ray-tracing calculation may use units $A=1$. Then
\begin{equation}
 \bar{r}=\frac{r}{A},
 \qquad
 \bar{B}=\frac{B}{A},
 \qquad
 \bar{X}=\frac{X}{A}.
 \label{eq:dimensionless_vars}
\end{equation}

The metric functions become
\begin{equation}
 {\cal H}(\bar{r})
 =
 (1+\alpha)-\frac{1+\bar{B}^2}{\bar{r}^2},
 \qquad
 {\cal G}(\bar{r})
 =
 1-\frac{1}{(1+\alpha)\bar{r}^2},
 \label{eq:dimensionless_HG}
\end{equation}
and
\begin{equation}
 \gamma(\bar{r})
 =
 A\bar{r}
 \left[
 1+\frac{2\alpha(1+\bar{B})}{\bar{r}}
 \right]^{1/2}.
 \label{eq:dimensionless_gamma}
\end{equation}
Representative parameter values are
\begin{equation}
 \alpha=0,\;0.05,\;0.10,\;0.15,
 \qquad
 B/A=0,\;0.1,\;0.2,\;0.3 .
 \label{eq:fiducial_values}
\end{equation}

A convergence test should vary both the number of screen points and the radial integration resolution. The flux asymmetry is usually the slowest observable to converge because it depends on the tails of the profile. The redshift observables are often more stable because they are weighted by the same intensity distribution used in the denominator.

\begin{figure}[tbhp]
\centering
\includegraphics[width=0.98\linewidth]{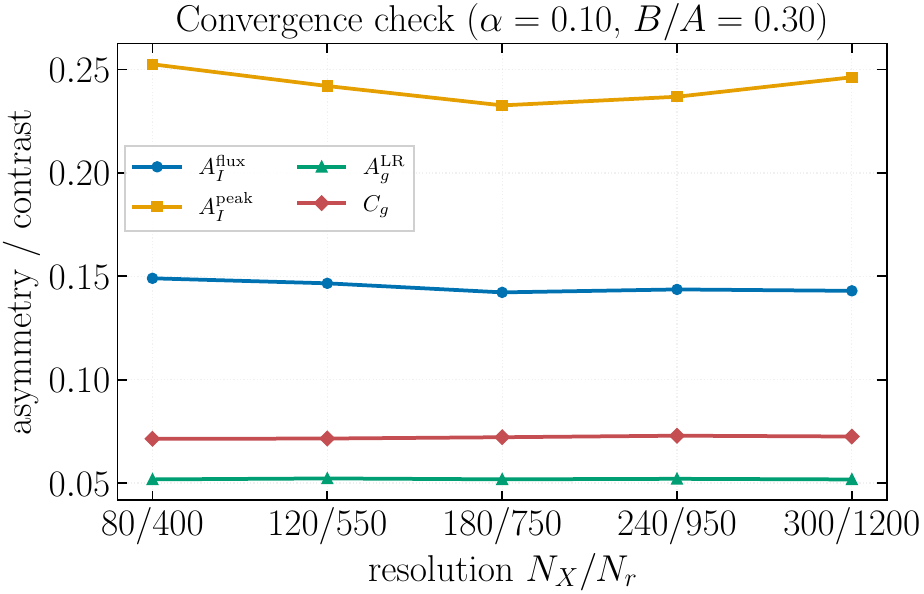}
\caption{Convergence check for the transfer observables at $\alpha=0.10$ and $B/A=0.30$.  The resolution label $N_X/N_r$ denotes the ratio of the number of screen points to the number of radial integration points.  The diagnostic hierarchy is stable at the level required for the qualitative transfer analysis, with peak and flux observables showing the strongest sensitivity to finite resolution.}
\label{fig:lv_convergence}
\end{figure}

Figure~\ref{fig:lv_convergence} shows a representative convergence test for $\alpha=0.10$ and $B/A=0.30$.  The labels $N_X/N_r$ denote the number of screen points and radial integration points used in the transfer calculation.  As in the Lorentz-symmetric draining-bathtub case, the integrated flux asymmetry is comparatively sensitive to the profile tails and detector smoothing, whereas the redshift quantities are more stable once the same intensity weight is used in the numerator and denominator.

\subsection{Consistency checks}
\label{subsec:consistency_checks}

We performed several analytic and numerical checks to verify the internal consistency of the results.  First, in the Lorentz-symmetric limit $\alpha=0$, the critical interval reduces to the standard rotating draining-bathtub result,
\begin{align}
 \frac{X_c^\pm}{A}
 &=-2\frac{B}{A}\pm2\sqrt{1+\frac{B^2}{A^2}},
 \label{eq:consistency_alpha_zero_branches}\\
 X_{\rm mid}^{(0)}&=-2B,
 \qquad
 \Delta X_0=4A\sqrt{1+\frac{B^2}{A^2}} .
 \label{eq:consistency_alpha_zero}
\end{align}
These results are recovered analytically by setting $\alpha=0$ in Eqs.~\eqref{eq:critical_conditions} and \eqref{eq:bc_implicit}, which reduces $\gamma^2(r)\to r^2$ and ${\cal H}(r)\to 1-(A^2+B^2)/r^2$, and solving the resulting quartic for the critical radius. Second, in the nonrotating Lorentz-violating case, the two branches are symmetric, and Eq.~\eqref{eq:static_width} follows from a direct expansion of Eqs.~\eqref{eq:critical_conditions} and \eqref{eq:bc_implicit}.  Third, expanding the exact critical conditions at small $\alpha$ and $\beta=B/A$ gives the mixed correction in Eq.~\eqref{eq:weak_centroid} and the width correction in Eq.~\eqref{eq:weak_width}.  These limits were used as benchmarks for the numerical critical points shown in Fig.~\ref{fig:lv_shadow_interval}.

A useful branch-by-branch expansion of the physical critical screen coordinates is
\begin{align}
 \frac{X_c^+}{A}
 &\simeq
 2-2\beta+(\sqrt{2}-1)\alpha
 +\left(1+\frac{3\sqrt{2}}{2}\right)\alpha\beta,
 \label{eq:Xc_plus_expansion}\\
 \frac{X_c^-}{A}
 &\simeq
 -2-2\beta-(\sqrt{2}-1)\alpha
 +\left(1-\frac{\sqrt{2}}{2}\right)\alpha\beta .
 \label{eq:Xc_minus_expansion}
\end{align}
Equations~\eqref{eq:Xc_plus_expansion} and \eqref{eq:Xc_minus_expansion} reproduce Eqs.~\eqref{eq:weak_centroid} and \eqref{eq:weak_width} after taking the midpoint and difference.  They also show that the mixed Lorentz-rotation correction is branch-dependent: the coefficient of $\alpha\beta$ is larger on the positive screen boundary than on the negative one. This asymmetry is not simply a statement about the coordinate radius of the circular orbit; rather, it results from the combined branch dependence of the critical solution, the circulation term, the Lorentz-deformed circumference function, and the asymptotic normalization $X=\sqrt{1+\alpha}\,b$.

The transfer calculation was checked by requiring that $I_{\rm obs}^{\rm det}(X)$ remain nonnegative, that the redshift factor be finite on the emitting segments, and that the critical roots satisfy both $\mathcal{R}(r_c;b_c)=0$ and $\partial_r\mathcal{R}(r_c;b_c)=0$ to numerical precision.  The convergence test in Fig.~\ref{fig:lv_convergence} then checks the full pipeline: critical-interval classification, radial integration, redshift weighting, detector convolution and extraction of the differential observables.  These tests support the qualitative conclusion that $\alpha$ primarily changes the global capture/redshift scale, while $B$ is responsible for the dominant left-right branch imbalance.

\section{Conclusion}
\label{sec:conclusion}

We have developed a transfer-observable framework for the rotating acoustic black hole with Lorentz symmetry violation. Starting from the Lorentz-violating draining-bathtub metric, we derived the null-ray equations, the critical-impact-parameter conditions, the acoustic redshift factor, and the observed-intensity transfer prescription for thin rings and extended disks.

The analysis shows that Lorentz violation and rotation affect the acoustic image in distinct but coupled ways. The Lorentz-breaking parameter $\alpha$ modifies the global capture scale, the effective circumference function, the asymptotic redshift normalization, and the shadow width. The circulation parameter $B$ shifts the shadow centroid and generates branch-dependent redshift and flux asymmetries. Mixed $\alpha B$ terms alter the calibration of rotational observables, showing that the two parameters should be inferred jointly rather than independently.

The most robust observables naturally form a hierarchy. The shadow width is the leading geometric probe of Lorentz-violating broadening, while the shadow centroid is the leading geometric probe of rotation. The left-right acoustic-redshift asymmetry tests the branch-dependent frequency transfer, and the branch-integrated flux asymmetry measures the corresponding brightness imbalance. The peak asymmetry is useful but more sensitive to caustics, finite source width, and detector resolution. The global redshift contrast measures the radial spread of acoustic redshifts across the source and should not, by itself, be interpreted as a unique signature of rotation.

Several structural points were clarified. The screen coordinate $X=\sqrt{1+\alpha}\,b$ is the natural observable coordinate for an asymptotic static detector because the factor $\sqrt{1+\alpha}$ follows from the modified asymptotic sound speed. The physically regular branch considered here is $\alpha\geq0$, for which the effective circumference function remains positive throughout the exterior region; for sufficiently negative $\alpha$, the circumference function can vanish outside the horizon, and the effective geometry breaks down for exterior ray propagation. The branch-by-branch expansion also shows why the two critical boundaries respond differently to mixed Lorentz-rotation corrections: the asymmetry results from the combined branch dependence of the critical solution, the circulation term, the Lorentz-deformed circumference function, and the asymptotic normalization.

The two-dimensional screen analysis shows that the synthetic representation of the $(2+1)$-dimensional acoustic shadow is a vertical strip in the $(b_x,b_y)$ plane, rather than a circular disk. This is the natural visualization of the one-dimensional capture interval of the draining-bathtub geometry. When a nonaxisymmetric source is included, the observer-azimuth map $I(X,\phi_{\rm obs})$ develops a sinusoidal modulation whose amplitude depends on the source compactness: extended disks suppress this modulation, while thin annuli enhance it.

This work extends absorption and quasinormal-mode studies of Lorentz-violating acoustic black holes to the image-transfer level. Future developments should include dispersive acoustic rays, viscous damping, finite-depth corrections, nonaxisymmetric source reconstruction, and direct comparison with numerical wave simulations. A genuinely circular acoustic shadow would require a different, spherically symmetric $(3+1)$-dimensional acoustic geometry and is left for future work.

\section*{Data availability}

No experimental data are used in this theoretical analysis. Numerical data generated from the ray-tracing procedure described here can be produced from Eqs.~\eqref{eq:dphidr}, \eqref{eq:gac_general}, and \eqref{eq:disk_transfer}.

\nocite{*}
\bibliographystyle{apsrev4-2}
\bibliography{references}

@article{Unruh1981,
  author = {Unruh, W. G.},
  title = {Experimental black-hole evaporation?},
  journal = {Phys. Rev. Lett.},
  volume = {46},
  pages = {1351--1353},
  year = {1981},
  doi = {10.1103/PhysRevLett.46.1351}
}

@article{Visser1998,
  author = {Visser, Matt},
  title = {Acoustic black holes: horizons, ergospheres and Hawking radiation},
  journal = {Class. Quantum Grav.},
  volume = {15},
  pages = {1767--1791},
  year = {1998},
  doi = {10.1088/0264-9381/15/6/024},
  eprint = {gr-qc/9712010},
  archivePrefix = {arXiv}
}

@article{Barcelo2011,
  author = {Barcelo, Carlos and Liberati, Stefano and Visser, Matt},
  title = {Analogue Gravity},
  journal = {Living Rev. Relativ.},
  volume = {14},
  pages = {3},
  year = {2011},
  doi = {10.12942/lrr-2011-3},
  eprint = {gr-qc/0505065},
  archivePrefix = {arXiv}
}

@article{Novello2002,
  author = {Novello, M. and Visser, M. and Volovik, G. E.},
  title = {Artificial black holes},
  journal = {World Scientific},
  year = {2002}
}

@article{Garay2000,
  author = {Garay, L. J. and Anglin, J. R. and Cirac, J. I. and Zoller, P.},
  title = {Sonic analog of gravitational black holes in Bose-Einstein condensates},
  journal = {Phys. Rev. Lett.},
  volume = {85},
  pages = {4643--4647},
  year = {2000},
  doi = {10.1103/PhysRevLett.85.4643},
  eprint = {gr-qc/0002015},
  archivePrefix = {arXiv}
}

@article{Philbin2008,
  author = {Philbin, T. G. and Kuklewicz, C. and Robertson, S. and Hill, S. and Konig, F. and Leonhardt, U.},
  title = {Fiber-optical analog of the event horizon},
  journal = {Science},
  volume = {319},
  pages = {1367--1370},
  year = {2008},
  doi = {10.1126/science.1153625}
}

@article{Lahav2010,
  author = {Lahav, O. and Itah, A. and Blumkin, A. and Gordon, C. and Rinott, S. and Zayats, A. and Steinhauer, J.},
  title = {Realization of a sonic black hole analog in a Bose-Einstein condensate},
  journal = {Phys. Rev. Lett.},
  volume = {105},
  pages = {240401},
  year = {2010},
  doi = {10.1103/PhysRevLett.105.240401},
  eprint = {0906.1337},
  archivePrefix = {arXiv}
}

@article{Weinfurtner2011,
  author = {Weinfurtner, S. and Tedford, E. W. and Penrice, M. C. J. and Unruh, W. G. and Lawrence, G. A.},
  title = {Measurement of stimulated Hawking emission in an analogue system},
  journal = {Phys. Rev. Lett.},
  volume = {106},
  pages = {021302},
  year = {2011},
  doi = {10.1103/PhysRevLett.106.021302},
  eprint = {1008.1911},
  archivePrefix = {arXiv}
}

@article{Steinhauer2016,
  author = {Steinhauer, Jeff},
  title = {Observation of quantum Hawking radiation and its entanglement in an analogue black hole},
  journal = {Nature Phys.},
  volume = {12},
  pages = {959--965},
  year = {2016},
  doi = {10.1038/nphys3863}
}

@article{Torres2017,
  author = {Torres, Theo and Patrick, Sam and Coutant, Antonin and Richartz, Mauricio and Tedford, Edmund W. and Weinfurtner, Silke},
  title = {Observation of superradiance in a vortex flow},
  journal = {Nature Phys.},
  volume = {13},
  pages = {833--836},
  year = {2017},
  doi = {10.1038/nphys4151},
  eprint = {1612.06180},
  archivePrefix = {arXiv}
}

@article{Patrick2018,
  author = {Patrick, Sam and Coutant, Antonin and Richartz, Mauricio and Weinfurtner, Silke},
  title = {Black-hole quasibound states from a draining bathtub vortex flow},
  journal = {Phys. Rev. Lett.},
  volume = {121},
  pages = {061101},
  year = {2018},
  doi = {10.1103/PhysRevLett.121.061101},
  eprint = {1801.08473},
  archivePrefix = {arXiv}
}

@article{Munoz2019,
  author = {Munoz de Nova, J. R. and Golubkov, K. and Kolobov, V. I. and Steinhauer, J.},
  title = {Observation of thermal Hawking radiation and its temperature in an analogue black hole},
  journal = {Nature},
  volume = {569},
  pages = {688--691},
  year = {2019},
  doi = {10.1038/s41586-019-1241-0}
}

@article{Kolobov2021,
  author = {Kolobov, V. I. and Golubkov, K. and Munoz de Nova, J. R. and Steinhauer, J.},
  title = {Observation of stationary spontaneous Hawking radiation and the time evolution of an analogue black hole},
  journal = {Nature Phys.},
  volume = {17},
  pages = {362--367},
  year = {2021},
  doi = {10.1038/s41567-020-01076-0}
}

@article{Basak2003,
  author = {Basak, Soumen and Majumdar, Parthasarathi},
  title = {Superresonance from a rotating acoustic black hole},
  journal = {Class. Quantum Grav.},
  volume = {20},
  pages = {3907--3914},
  year = {2003},
  doi = {10.1088/0264-9381/20/18/308},
  eprint = {gr-qc/0203059},
  archivePrefix = {arXiv}
}

@article{Berti2004,
  author = {Berti, Emanuele and Cardoso, Vitor and Lemos, Jose P. S.},
  title = {Quasinormal modes and classical wave propagation in analogue black holes},
  journal = {Phys. Rev. D},
  volume = {70},
  pages = {124006},
  year = {2004},
  doi = {10.1103/PhysRevD.70.124006},
  eprint = {gr-qc/0408099},
  archivePrefix = {arXiv}
}

@article{Cardoso2004,
  author = {Cardoso, Vitor and Lemos, Jose P. S. and Yoshida, Shijun},
  title = {Quasinormal modes of Schwarzschild black holes in four and higher dimensions},
  journal = {Phys. Rev. D},
  volume = {69},
  pages = {044004},
  year = {2004},
  doi = {10.1103/PhysRevD.69.044004},
  eprint = {gr-qc/0309112},
  archivePrefix = {arXiv}
}

@article{Dolan2009,
  author = {Dolan, Sam R.},
  title = {Scattering and absorption of gravitational plane waves by rotating black holes},
  journal = {Class. Quantum Grav.},
  volume = {25},
  pages = {235002},
  year = {2008},
  doi = {10.1088/0264-9381/25/23/235002},
  eprint = {0801.3805},
  archivePrefix = {arXiv}
}

@article{Oliveira2010,
  author = {Oliveira, Ednilton S. and Dolan, Sam R. and Crispino, Luis C. B.},
  title = {Absorption of planar waves in a draining bathtub},
  journal = {Phys. Rev. D},
  volume = {81},
  pages = {124013},
  year = {2010},
  doi = {10.1103/PhysRevD.81.124013},
  eprint = {1005.4038},
  archivePrefix = {arXiv}
}

@article{Dolan2012,
  author = {Dolan, Sam R. and Oliveira, Leandro A. and Crispino, Luis C. B.},
  title = {Scattering by a draining bathtub vortex},
  journal = {Phys. Rev. D},
  volume = {85},
  pages = {044031},
  year = {2012},
  doi = {10.1103/PhysRevD.85.044031},
  eprint = {1105.1795},
  archivePrefix = {arXiv}
}

@article{Dolan2013,
  author = {Dolan, Sam R. and Oliveira, Leandro A. and Crispino, Luis C. B.},
  title = {Resonances of a rotating black hole analogue},
  journal = {Phys. Rev. D},
  volume = {87},
  pages = {124026},
  year = {2013},
  doi = {10.1103/PhysRevD.87.124026},
  eprint = {1211.3751},
  archivePrefix = {arXiv}
}

@article{Berti2009,
  author = {Berti, Emanuele and Cardoso, Vitor and Starinets, Andrei O.},
  title = {Quasinormal modes of black holes and black branes},
  journal = {Class. Quantum Grav.},
  volume = {26},
  pages = {163001},
  year = {2009},
  doi = {10.1088/0264-9381/26/16/163001},
  eprint = {0905.2975},
  archivePrefix = {arXiv}
}

@article{Colladay1997,
  author = {Colladay, Don and Kostelecky, V. Alan},
  title = {CPT violation and the standard model},
  journal = {Phys. Rev. D},
  volume = {55},
  pages = {6760--6774},
  year = {1997},
  doi = {10.1103/PhysRevD.55.6760},
  eprint = {hep-ph/9703464},
  archivePrefix = {arXiv}
}

@article{Colladay1998,
  author = {Colladay, Don and Kostelecky, V. Alan},
  title = {Lorentz-violating extension of the standard model},
  journal = {Phys. Rev. D},
  volume = {58},
  pages = {116002},
  year = {1998},
  doi = {10.1103/PhysRevD.58.116002},
  eprint = {hep-ph/9809521},
  archivePrefix = {arXiv}
}

@article{Kostelecky2004,
  author = {Kostelecky, V. Alan},
  title = {Gravity, Lorentz violation, and the standard model},
  journal = {Phys. Rev. D},
  volume = {69},
  pages = {105009},
  year = {2004},
  doi = {10.1103/PhysRevD.69.105009},
  eprint = {hep-th/0312310},
  archivePrefix = {arXiv}
}

@article{Mattingly2005,
  author = {Mattingly, David},
  title = {Modern tests of Lorentz invariance},
  journal = {Living Rev. Relativ.},
  volume = {8},
  pages = {5},
  year = {2005},
  doi = {10.12942/lrr-2005-5},
  eprint = {gr-qc/0502097},
  archivePrefix = {arXiv}
}

@article{Liberati2013,
  author = {Liberati, Stefano},
  title = {Tests of Lorentz invariance: a 2013 update},
  journal = {Class. Quantum Grav.},
  volume = {30},
  pages = {133001},
  year = {2013},
  doi = {10.1088/0264-9381/30/13/133001},
  eprint = {1304.5795},
  archivePrefix = {arXiv}
}

@article{Anacleto2011,
  author = {Anacleto, M. A. and Brito, F. A. and Passos, E.},
  title = {Superresonance effect from a rotating acoustic black hole and Lorentz symmetry breaking},
  journal = {Phys. Lett. B},
  volume = {703},
  pages = {609--613},
  year = {2011},
  doi = {10.1016/j.physletb.2011.08.056},
  eprint = {1101.2891},
  archivePrefix = {arXiv}
}

@article{Anacleto2012,
  author = {Anacleto, M. A. and Brito, F. A. and Passos, E.},
  title = {Analogue Aharonov-Bohm effect in a Lorentz-violating background},
  journal = {Phys. Rev. D},
  volume = {86},
  pages = {125015},
  year = {2012},
  doi = {10.1103/PhysRevD.86.125015},
  eprint = {1208.2615},
  archivePrefix = {arXiv}
}

@article{Anacleto2019,
  author = {Anacleto, M. A. and Brito, F. A. and Passos, E.},
  title = {Quantum-corrected acoustic black holes in Lorentz-violating Abelian Higgs model},
  journal = {Phys. Lett. B},
  volume = {788},
  pages = {231--237},
  year = {2019},
  doi = {10.1016/j.physletb.2018.11.047}
}

@article{Campos2026,
  author = {Campos, J. A. V. and Anacleto, M. A. and Brito, F. A. and Passos, E. and Queiroz, Amilcar R.},
  title = {Absorption and quasinormal modes by rotating acoustic black holes in Lorentz-violating background},
  journal = {arXiv preprint},
  year = {2026},
  eprint = {2604.06137},
  archivePrefix = {arXiv},
  primaryClass = {gr-qc}
}

@article{Ahmed2026Transfer,
  author = {Ahmed, Faizuddin and Al-Badawi, Ahmad and Belchior, Fernando M. and Silva, Edilberto O.},
  title = {Transfer observables of rotating acoustic black holes from ray tracing: shadow centroid, redshift asymmetry and flux imbalance},
  journal = {arXiv preprint},
  year = {2026},
  eprint = {2605.20354},
  archivePrefix = {arXiv},
  primaryClass = {gr-qc}
}

@article{Synge1966,
  author = {Synge, J. L.},
  title = {The escape of photons from gravitationally intense stars},
  journal = {Mon. Not. R. Astron. Soc.},
  volume = {131},
  pages = {463--466},
  year = {1966},
  doi = {10.1093/mnras/131.3.463}
}

@incollection{Bardeen1973,
  author = {Bardeen, James M.},
  title = {Timelike and null geodesics in the Kerr metric},
  booktitle = {Black Holes},
  editor = {DeWitt, C. and DeWitt, B. S.},
  publisher = {Gordon and Breach},
  address = {New York},
  pages = {215--239},
  year = {1973}
}

@article{Cunningham1973,
  author = {Cunningham, C. T. and Bardeen, J. M.},
  title = {The optical appearance of a star orbiting an extreme Kerr black hole},
  journal = {Astrophys. J.},
  volume = {183},
  pages = {237--264},
  year = {1973},
  doi = {10.1086/152223}
}

@article{Luminet1979,
  author = {Luminet, J.-P.},
  title = {Image of a spherical black hole with thin accretion disk},
  journal = {Astron. Astrophys.},
  volume = {75},
  pages = {228--235},
  year = {1979}
}

@article{Falcke2000,
  author = {Falcke, Heino and Melia, Fulvio and Agol, Eric},
  title = {Viewing the shadow of the black hole at the Galactic Center},
  journal = {Astrophys. J. Lett.},
  volume = {528},
  pages = {L13--L16},
  year = {2000},
  doi = {10.1086/312423},
  eprint = {astro-ph/9912263},
  archivePrefix = {arXiv}
}

@article{Cunha2018,
  author = {Cunha, Pedro V. P. and Herdeiro, Carlos A. R.},
  title = {Shadows and strong gravitational lensing: a brief review},
  journal = {Gen. Relativ. Gravit.},
  volume = {50},
  pages = {42},
  year = {2018},
  doi = {10.1007/s10714-018-2361-9},
  eprint = {1801.00860},
  archivePrefix = {arXiv}
}

@article{Gralla2019,
  author = {Gralla, Samuel E. and Holz, Daniel E. and Wald, Robert M.},
  title = {Black hole shadows, photon rings, and lensing rings},
  journal = {Phys. Rev. D},
  volume = {100},
  pages = {024018},
  year = {2019},
  doi = {10.1103/PhysRevD.100.024018},
  eprint = {1906.00873},
  archivePrefix = {arXiv}
}

@article{Gralla2020,
  author = {Gralla, Samuel E. and Lupsasca, Alexandru},
  title = {Lensing by Kerr black holes},
  journal = {Phys. Rev. D},
  volume = {101},
  pages = {044031},
  year = {2020},
  doi = {10.1103/PhysRevD.101.044031},
  eprint = {1910.12873},
  archivePrefix = {arXiv}
}

@article{Johnson2020,
  author = {Johnson, Michael D. and others},
  title = {Universal interferometric signatures of a black hole's photon ring},
  journal = {Sci. Adv.},
  volume = {6},
  pages = {eaaz1310},
  year = {2020},
  doi = {10.1126/sciadv.aaz1310},
  eprint = {1907.04329},
  archivePrefix = {arXiv}
}

@article{Perlick2022,
  author = {Perlick, Volker and Tsupko, Oleg Yu.},
  title = {Calculating black hole shadows: review of analytical studies},
  journal = {Phys. Rep.},
  volume = {947},
  pages = {1--39},
  year = {2022},
  doi = {10.1016/j.physrep.2021.10.004},
  eprint = {2105.07101},
  archivePrefix = {arXiv}
}

@article{EHT2019I,
  author = {{Event Horizon Telescope Collaboration}},
  title = {First M87 Event Horizon Telescope Results. I. The shadow of the supermassive black hole},
  journal = {Astrophys. J. Lett.},
  volume = {875},
  pages = {L1},
  year = {2019},
  doi = {10.3847/2041-8213/ab0ec7},
  eprint = {1906.11238},
  archivePrefix = {arXiv}
}

@article{EHT2019VI,
  author = {{Event Horizon Telescope Collaboration}},
  title = {First M87 Event Horizon Telescope Results. VI. The shadow and mass of the central black hole},
  journal = {Astrophys. J. Lett.},
  volume = {875},
  pages = {L6},
  year = {2019},
  doi = {10.3847/2041-8213/ab1141},
  eprint = {1906.11243},
  archivePrefix = {arXiv}
}

@article{EHT2022I,
  author = {{Event Horizon Telescope Collaboration}},
  title = {First Sagittarius A* Event Horizon Telescope Results. I. The shadow of the supermassive black hole in the center of the Milky Way},
  journal = {Astrophys. J. Lett.},
  volume = {930},
  pages = {L12},
  year = {2022},
  doi = {10.3847/2041-8213/ac6674}
}

@article{EHT2022VI,
  author = {{Event Horizon Telescope Collaboration}},
  title = {First Sagittarius A* Event Horizon Telescope Results. VI. Testing the black hole metric},
  journal = {Astrophys. J. Lett.},
  volume = {930},
  pages = {L17},
  year = {2022},
  doi = {10.3847/2041-8213/ac6756}
}

@article{Narayan2019,
  author = {Narayan, Ramesh and Johnson, Michael D. and Gammie, Charles F.},
  title = {The shadow of a spherically accreting black hole},
  journal = {Astrophys. J. Lett.},
  volume = {885},
  pages = {L33},
  year = {2019},
  doi = {10.3847/2041-8213/ab518c},
  eprint = {1910.02957},
  archivePrefix = {arXiv}
}

@article{Dexter2016,
  author = {Dexter, Jason},
  title = {A public code for general relativistic, polarised radiative transfer around spinning black holes},
  journal = {Mon. Not. R. Astron. Soc.},
  volume = {462},
  pages = {115--136},
  year = {2016},
  doi = {10.1093/mnras/stw1526},
  eprint = {1602.03184},
  archivePrefix = {arXiv}
}

@article{Lindquist1966,
  author = {Lindquist, Richard W.},
  title = {Relativistic transport theory},
  journal = {Ann. Phys.},
  volume = {37},
  pages = {487--518},
  year = {1966},
  doi = {10.1016/0003-4916(66)90207-7}
}

@book{Chandrasekhar1983,
  author = {Chandrasekhar, S.},
  title = {The Mathematical Theory of Black Holes},
  publisher = {Oxford University Press},
  address = {Oxford},
  year = {1983}
}

@article{FischerVisser2002,
  author = {Fischer, Uwe R. and Visser, Matt},
  title = {Riemannian Geometry of Irrotational Vortex Acoustics},
  journal = {Phys. Rev. Lett.},
  volume = {88},
  pages = {110201},
  year = {2002},
  doi = {10.1103/PhysRevLett.88.110201}
}

@article{FedichevFischer2004PRA,
  author = {Fedichev, Petr O. and Fischer, Uwe R.},
  title = {{``Cosmological'' quasiparticle production in harmonically trapped superfluid gases}},
  journal = {Phys. Rev. A},
  volume = {69},
  pages = {033602},
  year = {2004},
  doi = {10.1103/PhysRevA.69.033602},
  eprint = {cond-mat/0303063},
  archivePrefix = {arXiv}
}

@article{FischerSchutzhold2004,
  author = {Fischer, Uwe R. and Sch{\"u}tzhold, Ralf},
  title = {Quantum simulation of cosmic inflation in two-component Bose-Einstein condensates},
  journal = {Phys. Rev. A},
  volume = {70},
  pages = {063615},
  year = {2004},
  doi = {10.1103/PhysRevA.70.063615},
  eprint = {cond-mat/0406470},
  archivePrefix = {arXiv}
}

@article{FedichevFischer2003,
  author = {Fedichev, Petr O. and Fischer, Uwe R.},
  title = {Gibbons-Hawking Effect in the Sonic de Sitter Space-Time of an Expanding Bose-Einstein-Condensed Gas},
  journal = {Phys. Rev. Lett.},
  volume = {91},
  pages = {240407},
  year = {2003},
  doi = {10.1103/PhysRevLett.91.240407},
  eprint = {cond-mat/0304342},
  archivePrefix = {arXiv}
}

@article{FedichevFischer2004PRD,
  author = {Fedichev, Petr O. and Fischer, Uwe R.},
  title = {Observer dependence for the phonon content of the sound field living on the effective curved space-time background of a Bose-Einstein condensate},
  journal = {Phys. Rev. D},
  volume = {69},
  pages = {064021},
  year = {2004},
  doi = {10.1103/PhysRevD.69.064021},
  eprint = {cond-mat/0307200},
  archivePrefix = {arXiv}
}

@article{ChaFischer2017,
  author = {Ch{\"a}, Seok-Yeong and Fischer, Uwe R.},
  title = {Probing the Scale Invariance of the Inflationary Power Spectrum in Expanding Quasi-Two-Dimensional Dipolar Condensates},
  journal = {Phys. Rev. Lett.},
  volume = {118},
  pages = {130404},
  year = {2017},
  doi = {10.1103/PhysRevLett.118.130404},
  eprint = {1609.06155},
  archivePrefix = {arXiv}
}

@article{ChandranFischer2025,
  author = {Chandran, S. Mahesh and Fischer, Uwe R.},
  title = {Expansion-contraction duality breaking in a Planck-scale sensitive cosmological quantum simulator},
  journal = {Eur. Phys. J. C},
  volume = {85},
  pages = {1476},
  year = {2025},
  doi = {10.1140/epjc/s10052-025-15187-6},
  eprint = {2506.02719},
  archivePrefix = {arXiv}
}

@article{SchuetzholdUhlmannXuFischer2005,
  author = {Sch{\"u}tzhold, Ralf and Uhlmann, Michael and Xu, Yan and Fischer, Uwe R.},
  title = {Quantum backreaction in dilute Bose-Einstein condensates},
  journal = {Phys. Rev. D},
  volume = {72},
  pages = {105005},
  year = {2005},
  doi = {10.1103/PhysRevD.72.105005}
}

@article{BaakRibeiroFischer2022,
  author = {Baak, Sang-Shin and Ribeiro, Caio C. Holanda and Fischer, Uwe R.},
  title = {Number-conserving solution for dynamical quantum backreaction in a Bose-Einstein condensate},
  journal = {Phys. Rev. A},
  volume = {106},
  pages = {053319},
  year = {2022},
  doi = {10.1103/PhysRevA.106.053319},
  eprint = {2206.11317},
  archivePrefix = {arXiv}
}

@article{PalFischer2024,
  author = {Pal, Kunal and Fischer, Uwe R.},
  title = {Quantum nonlinear effects in the number-conserving analogue gravity of Bose-Einstein condensates},
  journal = {Phys. Rev. D},
  volume = {110},
  pages = {116022},
  year = {2024},
  doi = {10.1103/PhysRevD.110.116022},
  eprint = {2410.13596},
  archivePrefix = {arXiv}
}

@article{HuangZhangGuoChen2024,
  author = {Huang, Jiewei and Zhang, Zhenyu and Guo, Minyong and Chen, Bin},
  title = {Images and flares of geodesic hot spots around a Kerr black hole},
  journal = {Phys. Rev. D},
  volume = {109},
  pages = {124062},
  year = {2024},
  doi = {10.1103/PhysRevD.109.124062},
  eprint = {2402.16293},
  archivePrefix = {arXiv}
}

@article{HuangZhengGuoChen2024,
  author = {Huang, Jiewei and Zheng, Liheng and Guo, Minyong and Chen, Bin},
  title = {Coport: A New Public Code for Polarized Radiative Transfer in a Covariant Framework},
  journal = {arXiv e-prints},
  pages = {arXiv:2407.10431},
  year = {2024},
  eprint = {2407.10431},
  archivePrefix = {arXiv}
}

@article{ZhongHuYanGuoChen2021,
  author = {Zhong, Zhen and Hu, Zezhou and Yan, Haopeng and Guo, Minyong and Chen, Bin},
  title = {QED effects on Kerr black hole shadows immersed in uniform magnetic fields},
  journal = {Phys. Rev. D},
  volume = {104},
  pages = {104028},
  year = {2021},
  doi = {10.1103/PhysRevD.104.104028},
  eprint = {2108.06140},
  archivePrefix = {arXiv}
}

\end{document}